\documentclass[lettersize,journal]{IEEEtran}
\usepackage{amsmath,amsfonts}
\usepackage{algorithm}
\usepackage{algpseudocode}
\usepackage{array}
\usepackage{textcomp}
\usepackage{soul}
\usepackage{tabularx}
\usepackage{subcaption}
\usepackage{multirow}
\usepackage{makecell} 
\usepackage{romannum}
\usepackage{stfloats}
\usepackage{url}
\usepackage{xcolor}
\usepackage{verbatim}
\usepackage{graphicx}
\usepackage{cite}
\usepackage{ulem}

\newcommand{\algorithmicinitialize}{\textbf{Initialize}} % Text to display
\algnewcommand{\Initialize}{\State \algorithmicinitialize\ } % Line-numbered command
\newcommand{\algorithmicinput}{\textbf{Input:}}
\algnewcommand{\Input}{\State \algorithmicinput\ }
\newcommand{\algorithmicoutput}{\textbf{Output:}}
\algnewcommand{\Output}{\State \algorithmicoutput\ }

\begin{document}
\title{A Metaheuristic Framework for Optimized HAPS-Aided Localization in Urban Areas}

\author{Hongzhao Zheng,~\IEEEmembership{Member,~IEEE}, Mohamed Atia,~\IEEEmembership{Senior Member,~IEEE}, Halim Yanikomeroglu,~\IEEEmembership{Fellow,~IEEE}
        % <-this % stops a space
\thanks{This work was supported by NSERC Discovery Grant.}% <-this % stops a space
\thanks{H. Zheng is with the Embedded and Multi-sensor Systems Lab (EMSLab) and Carleton-NTN (Non-Terrestrial Networks) Lab, Department of Systems and Computer Engineering, Carleton University, Ottawa, ON, K1S 5B6, Canada (e-mail: hongzhaozheng@cmail.carleton.ca).}% <-this % stops a space
\thanks{M. Atia is with the Embedded and Multi-sensor Systems Lab (EMSLab), Department of Systems and Computer Engineering, Carleton University, Ottawa, ON, K1S 5B6, Canada (e-mail: Mohamed.atia@carleton.ca).}
\thanks{H. Yanikomeroglu is with Carleton-NTN (Non-Terrestrial Networks) Lab, Department of Systems and Computer Engineering, Carleton University, Ottawa, ON, K1S 5B6, Canada (e-mail: halim@sce.carleton.ca).}
}

% The paper headers
\markboth{}%
{Shell \MakeLowercase{\textit{et al.}}: A Sample Article Using IEEEtran.cls for IEEE Journals}

% Remember, if you use this you must call \IEEEpubidadjcol in the second
% column for its text to clear the IEEEpubid mark.

\maketitle

\begin{abstract}
High-altitude platform stations (HAPS), originally designed for communication services, can also provide structured signals of opportunity (SoOP) to augment the global navigation satellite system (GNSS). However, dense urban environments introduce severe blockage and non-line-of-sight (NLOS) conditions that undermine GNSS accuracy and render geometric placement metrics insufficient. To address this, we propose a metaheuristic framework for jointly optimizing the number and placement of HAPS under practical constraints by integrating high-fidelity 3D city models, ray-tracing, and multi-objective optimization to handle the discrete and highly non-convex design space. Three metaheuristic solutions based on distinct search principles are developed to efficiently explore the solution space, all demonstrating rapid convergence and consistently outperforming a greedy baseline, particularly in the low-to-moderate HAPS regime. For representative dense urban scenarios, we show that four HAPS are sufficient to satisfy an 18-m average 3D positioning error bound (PEB) threshold, while configurations with two to five HAPS achieve over 50\% reduction in mean and root mean square (RMS) PEB and up to 94\% and 87\% reduction in standard deviation and coefficient of variation (CV), respectively, compared to the satellite-only case. Diminishing returns are observed beyond six HAPS due to geometric redundancy, emphasizing the importance of optimized placement. The framework further demonstrates strong robustness and generalizability across diverse urban environments with varying building morphology and propagation conditions, establishing it as an effective and scalable solution for HAPS-assisted localization in realistic urban settings.
\end{abstract}


\section{Introduction}

\IEEEPARstart{H}{igh} altitude platform stations (HAPS), which are envisioned quasi-stationary aerial platforms operating in the stratosphere, have been extensively studied for their ability to support communication services~\cite{b32, b15}. Although HAPS are primarily designed for communication rather than navigation, their downlink signals with known timing and waveform characteristics can function as structured signals of opportunity (SoOP). Unlike conventional terrestrial SoOP such as WiFi, LTE, or 5G positioning reference signal (PRS), HAPS can operate at high elevation angles and can be equipped with satellite-grade atomic clocks, thereby mitigating key SoOP limitations such as poor geometry, severe multipath, and transmitter clock instability~\cite{b25,b30}. Owing to their high altitude (typically between 18 and 22 km~\cite{b15}) and favorable line-of-sight (LOS) conditions~\cite{b19,b20}, recent studies have shown that HAPS are capable of enhancing global navigation satellite system (GNSS)-based localization accuracy in both horizontal and vertical dimensions~\cite{b8, b9, b16}.

However, the complex urban landscapes and irregular building structures commonly found in city environments continue to present significant obstacles. Signal blockages remain prevalent, and some areas suffer from consistently poor localization performance due to persistent non-line-of-sight (NLOS) conditions. As a result, placement strategies based solely on geometric dilution of precision (GDOP) may fail to reflect these practical obstructions. While adding more strategically placed ranging sources can generally improve positioning accuracy, deploying a large number of HAPS can lead to increased system complexity and higher operational costs. Therefore, it is essential to adopt an optimization strategy that not only enhances overall localization performance but also limits the HAPS count.

Due to the discrete nature of the HAPS count and the environment-dependent objective function, the joint optimization of HAPS count and placement is a multi-modal multi-objective optimization problem (MMOP). As a result, traditional closed-form optimization techniques are inapplicable~\cite{b22}. While various convexification techniques exist~\cite{b33, b34}, they can only guarantee global optimality when the relaxation is exact, meaning that the optimal solution of the convexified problem also satisfies the constraints of the original non-convex problem and yields the same objective value. However, such exactness is not always assured; in many cases, the relaxation may not fully capture the original problem's feasible set. As a result, the solutions obtained may be suboptimal or even physically unrealizable in the original formulation~\cite{b10}.

As an alternative to traditional optimization techniques, metaheuristic methods have been widely explored and have shown strong practical performance in addressing MMOPs. Among them, genetic algorithms (GAs) are population-based evolutionary search methods inspired by natural selection. Unlike convex optimization techniques that require differentiability or problem convexification, GAs work directly on the original, potentially non-convex search space and do not rely on convexity or differentiability. While they do not guarantee global optimality, GAs are particularly well-suited for scenarios where high-quality, near-optimal solutions are acceptable and preferable to strict optimality under idealized assumptions. This makes GAs a strong candidate for solving MMOPs, where they have demonstrated robust performance across a variety of benchmark and real-world problems~\cite{b11, b12, b13, b28, b27}.

A widely adopted metaheuristic from the swarm intelligence family is multi-objective particle swarm optimization (MOPSO)~\cite{b23, b24}, which extends the traditional particle swarm optimization (PSO) framework to multi-objective settings. Similar to GAs, MOPSO is also a population-based approach that does not require convexity or gradient information, making it suitable for non-convex and non-differentiable problems. It draws inspiration from the collective behavior of social organisms, where each particle represents a potential solution and adapts its trajectory based on both personal and collective experiences.

Although both GAs and MOPSO are designed to tackle  multi-objective and non-convex optimization problems, their practical effectiveness often depends on the specific characteristics of the problem. For instance, for MMOPs with discrete variables, the resulting objective space becomes discontinuous. This discontinuity restricts the distribution of non-dominated solutions along the discrete dimension, thereby diminishing the diversity of the solution set. In MMOPs, the goal is not just to find a single optimal solution but to approximate the Pareto front by capturing a diverse set of trade-off solutions. Maintaining diversity is essential to avoid premature convergence, ensure comprehensive decision support, and account for structurally different configurations, especially when the decision space is discrete and non-convex. As a result, achieving a well-distributed set of trade-off solutions becomes more challenging, particularly for algorithms that rely on continuous variation to promote diversity. Consequently, many algorithms that are otherwise effective for MMOPs may become less suitable under such conditions~\cite{b38}.

In this paper, we propose a metaheuristic framework that jointly optimizes both the number and placement of HAPS to identify a minimal HAPS configuration that minimizes the average 3D position error bound (PEB) for specified regions of interest (ROI), while simultaneously minimizing the number of HAPS. Accordingly, each configuration is evaluated based on two objectives: average 3D PEB and HAPS count. Since satellite geometry repeats every sidereal day, the optimization is carried out over an entire sidereal day to ensure robustness for long-term deployment. It is important to note that, due to the metaheuristic nature of the framework, global optimality cannot be guaranteed; instead, the algorithms aim to efficiently identify high-quality, near-optimal configurations.

Considering the complex urban layouts and building geometries, a 3D city model combined with ray tracing is employed to assess the visibility between ranging sources (e.g. HAPS and satellites) and receivers. In general, the ROI can be a selection of areas within an urban environment, where localization accuracy is significantly degraded due to obstructions such as skyscrapers. Given the high computational cost of ray tracing, especially when considering all possible paths\cite{b17}, evaluating the PEB at every point is impractical. As a scalable alternative, signals are classified into LOS and NLOS paths. The average 3D PEB is then estimated over a set of ``representative locations'' selected to capture the challenging urban areas, which collectively define the ROI.

To provide a broad comparison of algorithmic behaviors and strengths in navigating the unique landscape of our two-objective, multi-modal optimization problem, we explore three metaheuristic algorithms with varied search principles: the adaptive special-crowding distance non-dominated genetic algorithm II (ASDNSGA-II), the non-dominated genetic algorithm III (NSGA-III), and MOPSO. 

ASDNSGA-II extends NSGA-II by introducing a special crowding distance (SCD)\footnote{In NSGA-II, \textit{crowding distance} is a metric that estimates how isolated a solution is by measuring the average distance to its neighbors in the objective space. Solutions with larger crowding distances are favored to maintain diversity along the Pareto front.} and an adaptive crossover mechanism~\cite{b27}. These additions are designed for multi-modal problems and help maintain diversity in both decision and objective spaces, which is essential in our discrete and highly non-convex search space. NSGA-III~\cite{b26}, originally developed for many-objective optimization, replaces the crowding-distance mechanism with a reference-point-based selection strategy based on perpendicular distance\footnote{\textit{Perpendicular distance} refers to the shortest orthogonal distance from a solution to the nearest reference line drawn from the origin through a predefined reference point.}. This systematic diversity-control mechanism remains beneficial even for two objectives, offering strong distribution guidance in challenging landscapes. This makes NSGA-III a suitable candidate for our setting, where maintaining diversity across discrete solutions is crucial. MOPSO incorporates Pareto dominance\footnote{A solution $\mathbf{A}$ \textit{Pareto-dominates} $\mathbf{B}$ if (1) it is no worse in all objectives and (2) strictly better in at least one objective.}, external archiving, and grid-based diversity preservation~\cite{b23}. These components guide the swarm effectively in both convergence and distribution. Given its relatively simple structure and demonstrated performance on benchmark problems with disconnected or multi-frontal Pareto sets, we adapt MOPSO’s core principles to our constrained, non-convex landscape.

% MOPSO incorporates key ideas such as Pareto dominance\footnote{A solution $\mathbf{A}$ is said to Pareto-dominates solution $\mathbf{B}$ if 1) $\mathbf{A}$ is no worse than $\mathbf{B}$ in all objectives, and 2) $\mathbf{A}$ is strictly better than $\mathbf{B}$ in at least one objective. The set of non-dominated solutions like $\mathbf{A}$ forms the Pareto front.~\cite{b31}.}, external archiving of non-dominated solutions, and a grid-based diversity preservation mechanism~\cite{b23}. These components work together to guide the swarm effectively in both convergence and distribution. Given the relatively simple structure of MOPSO and its demonstrated efficiency on diverse benchmark problems, including those with disconnected and multi-frontal Pareto fronts, we adapt its core principles to suit our constrained, non-convex optimization landscape.

% The fourth solution is derived from the multi-objective simulated annealing (MOSA) algorithm introduced in~\cite{b25}, a physics-inspired metaheuristic originally applied to the challenging task of hyperparameter optimization in convolutional neural networks (CNNs). Its underlying mechanism uses a dominance-based acceptance probability to explore complex, non-convex solution spaces while maintaining a balance between convergence and diversity. While our optimization problem differs significantly in structure and scale, the general characteristics of MOSA, particularly its capacity to navigate rugged search spaces, make it another potential candidate for adaptation within our framework.

To accommodate the discrete nature of the HAPS count and the structural constraints in our optimization model, all metaheuristic algorithms require tailored modifications. At a high level, we ensure (i) operators that can handle variable HAPS numbers, (ii) diversity measures suitable for discrete decision spaces, and (iii) consistent crossover, mutation, and selection mechanisms across algorithms. The detailed algorithm-specific adjustments for ASDNSGA-II, NSGA-III, and MOPSO are presented in Section~\Romannum{3} where we align these methods to operate fairly within our constrained, multi-modal search space.

The key contributions of this paper are summarized as follows:
\begin{enumerate}
    \item A novel optimization problem is proposed that aims to simultaneously optimize the number and placement of HAPS while satisfying practical constraints on PEB, HAPS altitude, and elevation mask.
    
    \item A high-fidelity simulation framework is developed which integrates satellite orbit propagation, a 3D city model, and ray-tracing-based LOS/NLOS classification to accurately evaluate the average 3D PEB for receivers distributed across the urban ROI.
    
    \item Three metaheuristic solutions with distinct search strategies are developed to address the discrete HAPS count and the non-convex search space, and their performance is validated against a fingerprinting-based baseline.
    
    \item To fairly assess the diversity of solutions with different HAPS counts in the decision space, we propose a decision-space crowding distance method based on the aggregated nearest-neighbor distance (ANND).
    
    \item For ASDNSGA-II, the original algorithm determines the crossover type based on the true Pareto set. However, since the true Pareto set is unknown in practice, we develop a heuristic crossover-type assignment strategy to approximate this behavior.
    
    \item For methodological consistency, we enhance solution diversity in NSGA-III by incorporating the same adaptive crossover strategy used in ASDNSGA-II. To achieve this, we introduce an adaptive method for setting the perpendicular-distance threshold.
    
\end{enumerate}

The remainder of the paper is organized as follows. Section~\Romannum{2} describes the system model and formulates the optimization problem. Section~\Romannum{3} outlines the proposed framework and details the adaptations made to ASDNSGA-II, NSGA-III, and MOPSO. Section~\Romannum{4} summarizes the simulation setup, including parameter configurations. Section~\Romannum{5} presents the simulation results, covering algorithmic and system performance, GDOP analysis, runtime and bottleneck analysis, and generalization across diverse city models. Finally, Section~\Romannum{6} concludes the paper.

% \textit{Notations}: Upper-case and lower-case boldface letters denote matrices and column vectors, respectively. Calligraphic uppercase letters denote sets. $(\cdot)^{\text{T}}$ denotes the transpose. $\mathbb{E} \left[ \cdot \right]$ is the expectation operator.

\section{Problem Formulation}
\begin{figure*}[t]  
    \centering
    \includegraphics[width=0.85\textwidth, height=0.35\textheight]{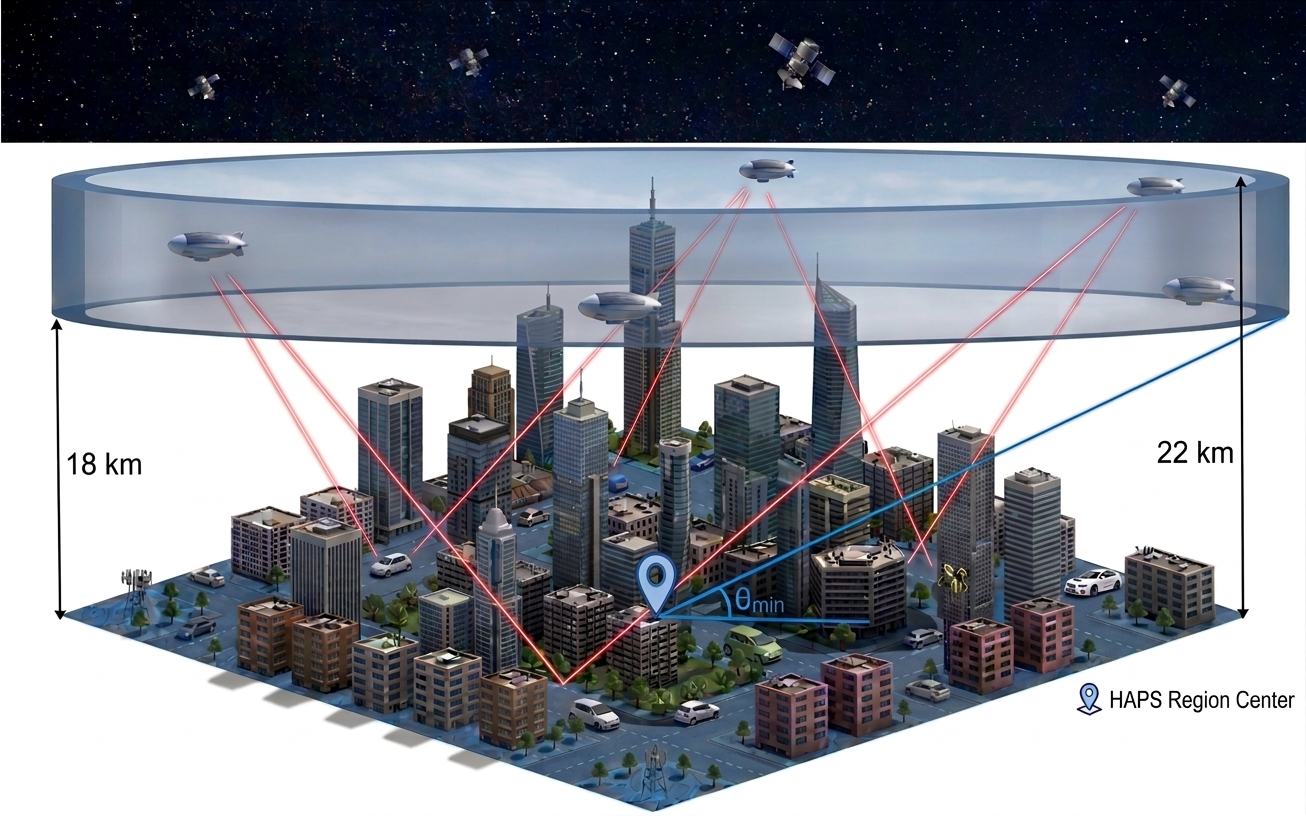}
    \caption{System model for HAPS-augmented GNSS localization in urban environments (HAPS are placed within a conical volume between 18 km and 22 km altitude while maintaining a minimum elevation angle $\theta_{\text{min}}$ relative to the region center).}
    \label{fig:system model}
\end{figure*}

The overall system model is illustrated in Fig.~\ref{fig:system model}. Let cars represent receivers located in areas with particularly poor signal conditions and low positioning accuracy. These areas are distributed throughout an urban environment, collectively defining the ROI. The receivers obtain ranging signals from both satellites and HAPS.

The ROI consists of $N_{\text{r}}$ receivers placed on the streets in a challenging urban environment. Let $\mathbf{P}^{\text{r}} = [\mathbf{p}^{\text{r}}_1, \ldots, \mathbf{p}^{\text{r}}_j, \ldots, \mathbf{p}^{\text{r}}_{N_{\text{r}}}]^{\text{T}}$ denote the matrix storing all receiver positions, where $\mathbf{p}^{\text{r}}_j = [\phi^{\text{r}}_j, \lambda^{\text{r}}_j, h^{\text{r}}_j]^{\text{T}}$ represents the latitude, longitude, and altitude of receiver $j$ for all $j\in\{ 1,2,\ldots,N_{\text{r}} \}$. Similarly, let $\mathbf{P}^{\text{s}} = [\mathbf{p}^{\text{s}}_1, \mathbf{p}^{\text{s}}_2, \ldots, \mathbf{p}^{\text{s}}_{N_{\text{s},j}}]^{\text{T}}$ denote the matrix of available\footnote{In this paper, \textit{available} means that a satellite or HAPS signal can be received by the receiver, regardless of whether the link is LOS or NLOS.} satellite positions for receiver $j$, where $N_{\text{s},j}$ may vary depending on the chosen elevation mask $\theta_{\text{min}}$. To ensure favorable geometry, the HAPS are constrained to maintain a minimum elevation angle relative to the region center $\mathbf{p}^{\text{c}}=[\phi^{\text{c}}, \lambda^{\text{c}}, h^{\text{c}}]$, and to remain within typically considered altitudes between 18 km and 22 km, thereby restricting their placement to a conical region in the sky. The elevation constraint is referenced to the region center, rather than individual receivers, because the ROI is geographically small compared to the HAPS altitude. From such distances, the angular variation across different receivers is negligible, making the center a representative reference point. This simplification also reduces computational cost by eliminating the need for receiver-specific constraints.

The primary objective is to identify the optimal configuration of HAPS that minimizes their number while maintaining an average 3D PEB below a predefined threshold. Furthermore, among all configurations with the same minimum HAPS count, the one with the lowest average 3D PEB is selected as the optimal solution. The optimization problem can be expressed as
\begin{subequations} \label{eq:opt1}
\begin{align}
    \underset{i \in \{1,2,\ldots,N_{\text{pop}}\}}{\text{min}} & \quad N_i \label{eq:opt1_obj} \\
    \text{s.t.} \quad 
    & \frac{1}{N_\text{t}N_\text{r}} \sum_{t=1}^{N_\text{t}}\sum_{j=1}^{N_\text{r}} \text{PEB}^{ijt} \leq \tau \label{eq:opt1_constraint}\\
    & N_i \in [N_{\text{min}}, N_{\text{max}}] \label{eq:opt1_constraint_number_of_haps}
\end{align}
\end{subequations}
\\[-4ex] % <<---- this reduces vertical space
\begin{subequations} \label{eq:opt2}
\begin{align}
    \underset{\mathbf{P}_l}{\text{min}} \quad 
    & \frac{1}{N_\text{t}N_\text{r}} \sum_{t=1}^{N_\text{t}}\sum_{j=1}^{N_{\text{r}}} \text{PEB}^{ijt} \label{eq:opt2_obj} \\
    \text{s.t.} \quad 
    & \mathbf{P}_l \in \mathcal{P}^\star \quad \forall l \in \{1, 2, \ldots, |\mathcal{P}^\star|\}\label{eq:opt2_constraint_cardinality}\\
    \quad & \theta_{k} \geq 10^{\circ} \quad \forall k \in \{1, 2, \ldots, N_i\} \label{eq:opt2_constraint_elevation}\\
    \quad & h_k \in [h_{\text{min}},h_{\text{max}}] \label{eq:opt2_constraint_altitude}
\end{align}
\end{subequations}

\noindent where $N_i$ denotes the number of HAPS selected for the $i^{\text{th}}$ candidate solution, $N_{\text{t}}$ denotes the number of snapshots, and $\tau$ is the predefined 3D PEB threshold. $N_{\text{min}}$ and $N_{\text{max}}$ represent the predefined minimum and maximum HAPS count, respectively. $N_{\text{pop}}$ denotes the population size, $\mathcal{P}^\star$ denotes the set of candidate solutions whose HAPS count is the least among all candidate solutions, and $|\mathcal{P}^\star|$ denotes the number of candidate solutions in $\mathcal{P}^\star$. $\mathbf{P}_l$ denotes the $l^{\text{th}}$ candidate solution in $\mathcal{P}^\star$, and $\theta_k$ denotes the elevation angle of the $k^{\text{th}}$ HAPS in a candidate solution. $\text{PEB}^{ijt}$ represents the position error bound associated with solution $i$, receiver $j$, and snapshot $t$. This can be computed as
\begin{equation}
\text{PEB}^{ijt} = \sqrt{\operatorname{Tr}\left( \mathbf{C}^{ijt}_{1:3,1:3}(\boldsymbol{\theta}^j) \right)}
\label{eq:peb}
\end{equation}

\noindent where $\mathbf{C}^{ijt}_{1:3,1:3}(\boldsymbol{\theta}^j)$ refers to the top-left $3 \times 3$ submatrix of the Cramér–Rao lower bound (CRLB) matrix associated with the $i^{\text{th}}$ candidate solution for receiver $j$ at snapshot $t$, evaluated at the parameter vector $\boldsymbol{\theta}^j=[x_{\text{r}}^j,y_{\text{r}}^j,z_{\text{r}}^j,cb_{\text{r}}^j]$. $[x_{\text{r}}^j, y_{\text{r}}^j,z_{\text{r}}^j]$ denote the x-, y-, and z-coordinates of the $j^{\text{th}}$ receiver in the Earth-centered Earth-fixed (ECEF) frame, $b_{\text{r}}^j$ denotes the $j^{\text{th}}$ receiver clock bias, and $c$ is the speed of light. Note that $\mathbf{C}^{ijt}(\boldsymbol{\theta}^j)$ accounts for contributions from both the satellites available to receiver $j$ and the HAPS included in the $i^{\text{th}}$ candidate solution.

The closed-form expression of the CRLB matrix is given by
\begin{equation} 
  \mathbf{C}(\boldsymbol{\theta}) = \mathbf{I}^{-1}(\boldsymbol{\theta})
  \label{eq:crlb}
\end{equation}
\begin{equation} 
\mathbf{I}(\boldsymbol{\theta}) = \mathbb{E} \left[ \left( \frac{\partial}{\partial \boldsymbol{\theta}} \log p(z; \boldsymbol{\theta}) \right)
\left( \frac{\partial}{\partial \boldsymbol{\theta}} \log p(z; \boldsymbol{\theta}) \right)^{\text{T}} \right]
\label{eq:fim_general}
\end{equation}

\noindent where $\mathbf{I}(\boldsymbol{\theta})$ is the Fisher information matrix (FIM) and $p(z; \boldsymbol{\theta})$ represents the likelihood function of observation $z$ given the parameter vector $\boldsymbol{\theta}$.

\begin{table*}[!t]
\caption{GMM parameters for pseudorange errors under LOS and NLOS conditions across all scenarios.}
\begin{center}
\renewcommand{\arraystretch}{1.3} % Vertical padding
\setlength{\tabcolsep}{10pt} % Horizontal padding
\begin{tabular}{c|c|c|c|c|c}
\Xhline{1pt}
\textbf{Scenario} & \textbf{Source} & \textbf{Condition} & \textbf{Mean(s) [m]} & \textbf{Std Dev(s) [m]} & \textbf{Weight(s)} \\
\Xhline{1pt}
\multirow{4}{*}{\textbf{Baseline}}
    & \multirow{2}{*}{Satellite} 
      & LOS  & 0                  & $10$                & 1 \\
    & & NLOS & $\{20, 40, 120\}$  & $\{15, 20, 50\}$    & $\{0.5, 0.4, 0.1\}$ \\
    \cline{2-6}
    & \multirow{2}{*}{HAPS}     
      & LOS  & 0                 & $7$              & 1 \\
    & & NLOS & $\{14, 28, 84\}$  & $\{10, 15, 35\}$ & $\{0.5, 0.4, 0.1\}$ \\
\hline
\multirow{4}{*}{\textbf{Extreme Reflection}}
    & \multirow{2}{*}{Satellite} 
      & LOS  & 0                  & $10$                & 1 \\
    & & NLOS & $\textcolor{red}{\{30, 60, 180\}}$  & $\{15, 20, 50\}$    & $\{0.5, 0.4, 0.1\}$ \\
    \cline{2-6}
    & \multirow{2}{*}{HAPS}     
      & LOS  & 0                 & $7$              & 1 \\
    & & NLOS & $\textcolor{red}{\{21, 42, 126\}}$  & $\{10, 15, 35\}$ & $\{0.5, 0.4, 0.1\}$ \\
\hline
\multirow{4}{*}{\textbf{Extreme NLOS}}
    & \multirow{2}{*}{Satellite} 
      & LOS  & 0                  & $10$                & 1 \\
    & & NLOS & $\{20, 40, 120\}$  & $\{15, 20, 50\}$    & $\textcolor{red}{\{0.2, 0.4, 0.4\}}$ \\
    \cline{2-6}
    & \multirow{2}{*}{HAPS}     
      & LOS  & 0                 & $7$              & 1 \\
    & & NLOS & $\{14, 28, 84\}$  & $\{10, 15, 35\}$ & $\textcolor{red}{\{0.2, 0.4, 0.4\}}$ \\
\hline
\multirow{4}{*}{\textbf{Low-End Receiver}}
    & \multirow{2}{*}{Satellite} 
      & LOS  & 0                  & $\textcolor{red}{15}$                & 1 \\
    & & NLOS & $\{20, 40, 120\}$  & $\textcolor{red}{\{22.5, 30, 75\}}$    & $\{0.5, 0.4, 0.1\}$ \\
    \cline{2-6}
    & \multirow{2}{*}{HAPS}     
      & LOS  & 0                 & $\textcolor{red}{10.5}$              & 1 \\
    & & NLOS & $\{14, 28, 84\}$  & $\textcolor{red}{\{15, 22.5, 52.5\}}$ & $\{0.5, 0.4, 0.1\}$ \\
\Xhline{1pt}
\end{tabular}
\label{tab:gmm_params}
\end{center}
\end{table*}

In urban environments, pseudorange errors are influenced by severe multipath and NLOS effects, in addition to atmospheric effects, ranging source position uncertainty, and receiver noise. As a result, these errors often exhibit complex, heavy-tailed, or multi-modal distributions and are more accurately modeled by Gaussian mixture models (GMMs)~\cite{b18}. In this work, for LOS conditions, both satellites and HAPS are modeled using zero-mean Gaussian distributions with  variances $\sigma^{\text{s}}_{\text{LOS}}$ and $\sigma^{\text{h}}_{\text{LOS}}$, respectively. For NLOS conditions, pseudorange errors are modeled using a GMM that exclude the LOS component, with each mixture characterized by non-zero means $\mu^{\text{s}}_{\text{NLOS}}$, $\mu^{\text{h}}_{\text{NLOS}}$, variances $\sigma^{\text{s}}_{\text{NLOS}}$, $\sigma^{\text{h}}_{\text{NLOS}}$, and associated weights $w$. 

To ensure a comprehensive evaluation, we assess the solution quality of the proposed metaheuristic framework under four sets of GMM parameters corresponding to different urban scenarios: 1) \textit{Baseline}: Satellite parameters are adopted from empirical error distributions observed in the Berlin Potsdamer Platz dataset, a representative dense urban environment~\cite{b35}. Since HAPS pseudorange errors are expected to be smaller than those of satellites~\cite{b8}, their means and variances are set to 70\% of the corresponding satellite values, while retaining the same mixture weights. 2) \textit{Extreme Reflection}: To account for stronger multipath-induced bias under severe reflection conditions, we increase the NLOS mean values by a factor of 1.5 relative to the baseline. 3) \textit{Extreme NLOS}: In this scenario, the mixture weights $w$ are modified to increase the likelihood of NLOS conditions. 4) \textit{Low-End Receiver}: To reflect degraded receiver quality and increased measurement noise, we increase the variances by a factor of 1.5 relative to the baseline. A summary of the GMM parameters for the different scenarios is provided in Table~\ref{tab:gmm_params}.

Under the assumed measurement model, the FIM is computed by aggregating the information contribution from each individual ranging source:
\begin{equation}
\mathbf{I}(\boldsymbol{\theta}) = \sum_{k=1}^{N} \psi_k \cdot \mathbf{h}_k^{\text{T}} \mathbf{h}_k
\label{eq:fim}
\end{equation}

\noindent where $\mathbf{h}_k$ denotes the $k$-th row of the design matrix $\mathbf{H}$ corresponding to a specific satellite or HAPS, and $\psi_k$ is the scalar information weight that quantifies the contribution of that link. For LOS links, $\psi_k$ is set as the inverse of the assumed Gaussian variance. For NLOS links, $\psi_k$ is obtained from the Fisher information of the GMM, computed through numerical integration over a discretized residual domain.

% For tractability, however, we assume that pseudorange noise follows a zero-mean Gaussian distribution, with standard deviation determined by the visibility condition. Specifically, $\sigma^{\text{h}}_{\text{LOS}}$ and $\sigma^{\text{s}}_{\text{LOS}}$ denote the standard deviations for HAPS and satellites under LOS conditions, while $\sigma^{\text{h}}_{\text{NLOS}}$ and $\sigma^{\text{s}}_{\text{NLOS}}$ represent the standard deviations under NLOS conditions. Under these assumptions, the FIM can be formulated as

% \begin{equation}
% \begin{aligned}
%   \mathbf{I}(\boldsymbol{\theta}, B) &= \mathbf{H}^\text{T} \mathbf{W}^{-1} \mathbf{H}, \\
%   \mathbf{H} &= 
%   \begin{bmatrix}
%     \mathbf{h}_1^{\text{T}} \\
%     \mathbf{h}_2^{\text{T}} \\
%     \vdots \\
%     \mathbf{h}_N^{\text{T}}
%   \end{bmatrix},
%   \quad
%   \mathbf{h}_i = 
%   \begin{bmatrix}
%     \frac{\mathbf{x} - \mathbf{p}_i^{\text{t}}}{\|\mathbf{x} - \mathbf{p}_i^{\text{t}}\|}\\
%     1
%   \end{bmatrix},\\
%   \mathbf{W} &= \text{diag}(\sigma_1^2, \sigma_2^2, \dots, \sigma_N^2)
% \end{aligned}
% \label{eq:fim}
% \end{equation}

% \noindent where $\mathbf{H}$ is known as the design matrix, $\mathbf{W}$ is a diagonal weight matrix containing the variances of the pseudorange noise for each ranging source, $\mathbf{x}$ and $\mathbf{p}_i^{\text{t}}$ denote the positions in the ECEF frame of the receiver and the $i^{\text{th}}$ ranging source, respectively. $\text{t}$ can be either a HAPS or a satellite, and $N$ is the sum the number of HAPS and satellites available to the receiver.

The formulated joint optimization problem evaluates the average PEB across the ROI and further averages it across multiple snapshots within one sidereal day, thereby capturing both variations in satellite geometry and differences in the receiver environment. This formulation is environment dependent, but it can be applied to other cities as long as an accurate 3D city model is available. However, jointly optimizing the number and placement of HAPS remains challenging for several reasons. First, the objective includes a cardinality term that penalizes the number of deployed HAPS, which is an inherently combinatorial and non-convex component. Second, the visibility conditions depend nonlinearly on HAPS placement relative to the environment. These factors lead to a highly non-convex, non-differentiable optimization landscape that motivates the use of evolutionary algorithms.

% Third, the information contributions $\psi_k$ depend on the LOS/NLOS conditions, which are determined by comparing the elevation angle of each ranging source with the corresponding skyline mask, after rounding to the nearest azimuth–elevation bin.

\section{Proposed Framework}
The proposed metaheuristic framework integrates satellite orbit propagation, HAPS location simulation, 3D city models, ray-tracing-based visibility analysis, and metaheuristic optimization algorithms. The 3D city model provides detailed building geometry, including geodetic coordinates and heights, and the road layout. The road layout is used to extract street-level nodes that serve as candidate receiver locations. To avoid repeated ray tracing operations in the main loop and thereby reduce computational complexity, a skyline mask\footnote{The \textit{skyline mask} specifies, for each azimuth direction, the \textit{skyline elevation}, which is the lowest elevation above the horizon at which the sky becomes unobstructed by nearby buildings.} is precomputed for each node using ray tracing based on the building geometry and node location. Street nodes exhibiting poor visibility are then identified based on their mean skyline elevation, and the region center is defined as the geodetic centroid of these selected poor-visibility nodes.

\begin{figure*}[t]
\centering
\includegraphics[width=0.8\textwidth]{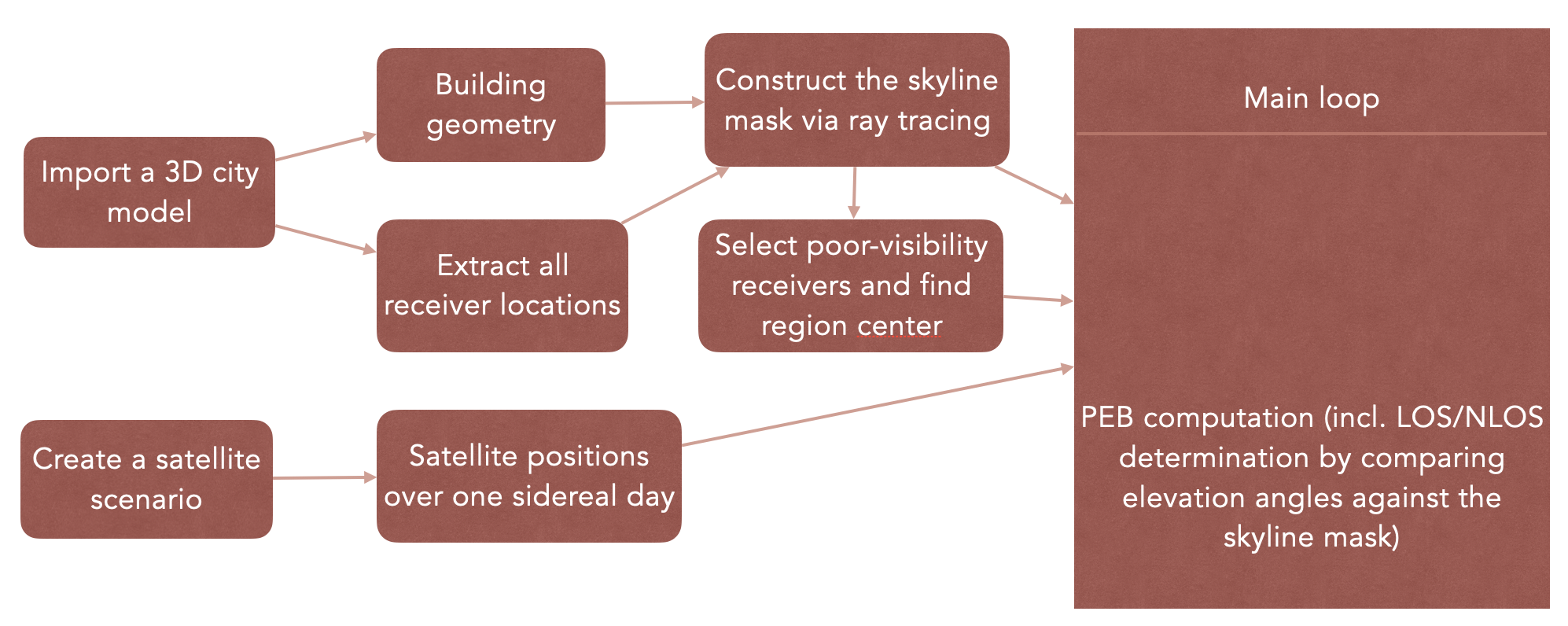}
\caption{Overview of the metaheuristic framework.}
\label{fig:framework workflow}
\end{figure*}

A high-fidelity satellite scenario that provides satellite positions throughout one sidereal day is another essential component of the framework. Using this scenario together with the predefined elevation mask, the set of available satellites for each receiver at any given timestamp is determined. The skyline masks, the identified poor-visibility receivers, the region-center coordinates, and the satellite positions are then provided as inputs to the main loop of the metaheuristic algorithms. Within this loop, PEB computation is performed for candidate HAPS configurations, including LOS/NLOS determination using the precomputed skyline masks. An overview of the proposed metaheuristic framework is shown in Fig.~\ref{fig:framework workflow}.

% To preserve the intrinsic characteristics of the problem and ensure a fair comparison, all three metaheuristic algorithms are carefully customized to optimize their performance while maintaining a high degree of methodological consistency across all approaches. In this section, the modifications made to each algorithm are described.

To accommodate the discrete nature of HAPS count, one of the decision variables, and the structural constraints inherent to our problem, each metaheuristic algorithm must be carefully modified. For instance, in both ASDNSGA-II and NSGA-III, the crossover and mutation operators are modified to allow variations in both the number and placement of HAPS. As a result, crossover can occur between individuals with different HAPS counts. To evaluate diversity in such a decision space, we develop a customized crowding distance measure based on the ANND. Furthermore, since the true Pareto set\footnote{In a multi-objective optimization problem, a solution is said to be \textit{Pareto optimal} if no other feasible solution exists that improves at least one objective without worsening another. The Pareto set is the set of all such Pareto-optimal solutions in the decision variable space~\cite{b31}.} is unknown in our problem, a heuristic crossover type assignment is developed as a replacement for the Pareto set proximity (PSP)-based crossover strategy adopted in~\cite{b27}.

To ensure methodological consistency across all approaches, we enhance solution diversity in NSGA-III by incorporating the same adaptive crossover strategy used in ASDNSGA-II. Following the core design of NSGA-III, the crowding distance in objective space is replaced by the perpendicular distance metric. We also integrate the heuristic crossover type assignment into NSGA-III. To enable its use, an adaptive perpendicular distance threshold method inspired by the distance dominance relation proposed in~\cite{b29} is introduced as part of the modified NSGA-III.

For MOPSO, a stratified archive update mechanism is developed to ensure compatibility during velocity\footnote{\textit{Velocity} in MOPSO is the update vector that moves a particle toward its personal best and a leader solution.} updates, where the number of HAPS must remain consistent across the particle's personal best, current state with previous velocity, and leader. In this scheme, non-dominated solutions are grouped according to their HAPS count. Furthermore, polynomial mutation is applied after the velocity update to maintain methodological consistency across all algorithms and to enhance solution diversity.

Overall, these modifications preserve the intrinsic characteristics of the problem while enabling a fair and meaningful comparison across all three metaheuristic approaches. The following subsections describe the algorithm-specific adjustments in detail.

\subsection{ASDNSGA-II Algorithm}
The core components of ASDNSGA-II include SCD computation, an improved binary tournament selection (IBTS) strategy, and an adaptive crossover switching mechanism that switches between simulated binary crossover (SBX) and blend crossover alpha (BLX-$\alpha$)~\cite{b27}. Unlike traditional crowding distance, which measures diversity only in the objective space, the SCD calculation considers both the decision and objective spaces to better preserve overall diversity. The IBTS prioritizes both the rank\footnote{For EAs, \textit{rank} refers to the non-domination level of a solution: Rank 1 includes all non-dominated individuals, Rank 2 includes those dominated only by Rank 1 individuals, and so on. This layered ranking guides selection and fitness assignment in the evolutionary process~\cite{b31}.} and the SCD, favoring candidate solutions, known as individuals, that not only have better ranks but also contribute to higher diversity. The adaptive crossover dynamically selects between SBX and BLX-$\alpha$ by comparing the PSP achieved by each crossover. 

\subsubsection{Crossover and Mutation}
As the decision variables include the count and spatial placement of HAPS, these aspects are explicitly incorporated into the BLX-$\alpha$ and SBX crossover operations, as well as the polynomial mutation operation, to ensure the proper evolution of solutions. Additionally, rounding is applied to the resulting HAPS count after crossover or mutation to retain its integer nature.

\subsubsection{Decision-Space Crowding Distance Based on ANND}
Given that individuals within the same Pareto front may have different numbers of HAPS, the standard crowding distance computation in the decision space is modified accordingly. Specifically, we adopt ANND to fairly assess the similarity between individuals. The ANND metric calculates all pairwise distances between two individuals and aggregates the minimum distances from each point in the individual with fewer HAPS to its nearest unmatched point in the individual with an equal or greater number of HAPS. This matching ensures that each HAPS in the larger individual is used at most once, thereby encouraging solution diversity.

\subsubsection{Heuristic Crossover Type Assignment}
Since the true Pareto set is unknown in our problem, the PSP-based crossover strategy exploited in~\cite{b27} cannot be directly applied. Instead, we adopt a heuristic approach to select the crossover type (SBX or BLX-$\alpha$) for each individual based on its rank $r$, the crowding distances in the decision space $d_{\text{dec}}$ and objective space $d_{\text{obj}}$, as well as the predefined thresholds $d_{\text{dec}}^{\text{th}}$ and $d_{\text{obj}}^{\text{th}}$. In particular, SBX is applied only when all the following conditions are met: $r=1$, $d_{\text{dec}} > d_{\text{dec}}^{\text{th}}$, and $d_{\text{obj}} > d_{\text{obj}}^{\text{th}}$. Otherwise, BLX-$\alpha$ is used.

\subsubsection{Algorithmic Workflow}
The algorithm begins by creating an initial population $\mathcal{P}_0$ consisting of $N_{\text{pop}}$ individuals, where each individual represents a candidate HAPS placement configuration containing $N_i$ HAPS, with $N_i \in [N_{\text{min}}, N_{\text{max}}]$. At the start of each generation, the objective matrix $\mathbf{F}_{\mathcal{P}}$, which has dimensions $N_{\text{pop}} \times 2$ and stores both the average 3D PEB and the HAPS count for each individual in $\mathcal{P}_{n-1}$, along with the offspring set $\mathcal{C}$ are initialized. For each individual in $\mathcal{P}_{n-1}$, the corresponding candidate solution is extracted as $\mathbf{P}_i$. The objectives of each individual are computed based on the receiver positions $\mathbf{P}^{\text{r}}$, satellite positions $\mathbf{P}^{\text{s}}$, and the skyline masks $\mathbf{M} = [\mathbf{m}_1(\phi), \mathbf{m}_2(\phi), \dots, \mathbf{m}_{N_{\text{r}}}(\phi)]^{\text{T}}$, where $\mathbf{m}_j(\phi)$ encodes the skyline elevation for receiver $j$ at azimuth $\phi$. The computation also incorporates an information weight vector $\boldsymbol{\psi}$, which is precomputed to capture the information contribution of both satellite and HAPS signals under LOS and NLOS conditions.

Subsequently, the fast non-dominated sorting (FNS) is performed on the objectives $\mathbf{F}_{\mathcal{P}}$ to obtain the Pareto front set $\mathcal{F}$ and the ranks $\mathbf{r}$ for all individuals. The elite solution $\mathbf{P}_n^\star$ is then identified through elite selection. Following this, the crowding distances in decision space $\mathbf{d}_{\text{dec}}$, objective space $\mathbf{d}_{\text{obj}}$, and the special crowding distances $\mathbf{d}_{\text{SCD}}$ are computed for all individuals, where both $\mathbf{d}_{\text{obj}}$ and $\mathbf{d}_{\text{SCD}}$ are calculated based on~\cite{b27}.

Parent selection is then performed using the IBTS from~\cite{b27}, resulting in $N_{\text{pop}}$ parents stored in $\mathcal{P}_n$. For each pair of parents, their ranks ($r_i$, $r_{i+1}$), decision-space crowding distances ($d_{\text{dec},i}$, $d_{\text{dec},{i+1}}$), and objective-space crowding distances ($d_{\text{obj},i}$, $d_{\text{obj},{i+1}}$) are retrieved. Adaptive crossover with the heuristic
crossover type assignment is then applied to $\mathbf{P}_i$ and $\mathbf{P}_{i+1}$. Subsequently, polynomial mutation is applied to introduce variations. The resulting children are added to the offspring set $\mathcal{C}_{n-1}$.

\begin{algorithm}[htbp]
\caption{Modified ASDNSGA-II for HAPS Placement and Count Joint Optimization}
\label{alg:modified asdnsga ii}
\begin{algorithmic}[1]
\Input $\mathbf{P}^{\text{r}}$, $\mathbf{P}^{\text{s}}$, $\mathbf{p}^{\text{c}}$, $\mathbf{M}$, $p_{\text{c}}$, $p_{\text{m}}$, $\boldsymbol{\psi}$, $\eta_{\text{c}}$, $\eta_{\text{m}}$, $N_{\text{pop}}$, $N_{\text{g}}$,
\Statex \hspace{2.6em} $N_{\text{min}}$, $N_{\text{max}}$, $\theta_{\text{min}}$, $\tau$.
\Output Optimal HAPS configuration $\mathbf{P}^{\star}_{N_{\text{g}}}$.
\Initialize $\mathcal{P}_0$.
\For{$n = 1 : N_{\text{g}}$}
    \Initialize $\mathbf{F}_{\mathcal{P}}$, $\mathcal{C}$.
    \For{$i = 1: N_{\text{pop}}$}
        \State $\mathbf{P}_i \gets \mathcal{P}_{n-1}$. 
        \State Compute $\mathbf{F}_{\mathcal{P}}$ based on $\mathbf{P}^{\text{r}}$, $\mathbf{P}^{\text{s}}$, $\mathbf{M}$, and $\boldsymbol{\psi}$.
        % \State Store objectives: average CRLB and number of HAPS
    \EndFor
    \State Compute $\mathcal{F}$ and $\mathbf{r}$ via FNS on $\mathbf{F}_{\mathcal{P}}$.
    \State Find the best solution $\mathbf{P}_n^\star$ via elite selection.
    \State Compute $\mathbf{d}_{\text{dec}}$, $\mathbf{d}_{\text{obj}}$, and $\mathbf{d}_{\text{SCD}}$ via SCD for all 
    \Statex \hspace{1.2em} individuals in $\mathcal{P}_{n-1}$ based on $\mathbf{F}_{\mathcal{P}}$ and $\mathcal{F}$.
    \State Select parents and store them in $\mathcal{P}_n$ via IBTS.
    \For{$i = 1 : 2 : N_{\text{pop}}$}
        \State $\mathbf{P}_i,\mathbf{P}_{i+1} \gets \mathcal{P}_n$.
        \State $r_i,r_{i+1} \gets \mathbf{r}$.
        \State $d_{\text{dec},i}, d_{\text{dec},{i+1}} \gets \mathbf{d}_{\text{dec}}$.
        \State $d_{\text{obj},i}, d_{\text{obj},{i+1}} \gets \mathbf{d}_{\text{obj}}$.
        \If{$rand \leq p_{\text{c}}$}
            \State {Apply the adaptive crossover with the heuristic 
            \Statex \hspace{4.2em} crossover type assignment on $\mathbf{P}_i$ and $\mathbf{P}_{i+1}$.}
        \EndIf
        \If{$rand \leq p_\text{m}$}
            \State Apply polynomial mutation on $\mathbf{P}_i$ and $\mathbf{P}_{i+1}$.
        \EndIf
        \State Add children $\mathbf{P}_i$ and $\mathbf{P}_{i+1}$ to offspring set $\mathcal{C}_{n-1}$.
    \EndFor
    \State Evaluate offspring objectives $\mathbf{F}_{\mathcal{C}}$.
    \State Combine population $\mathcal{P}_{n-1} \cup \mathcal{C}_{n-1}$.
    \State Combine objectives $\mathbf{F}_{\mathcal{P}} \cup \mathbf{F}_{\mathcal{C}}$.
    \State Get $\mathcal{P}_n$ via environmental selection on $\mathcal{P}_{n-1} \cup \mathcal{C}_{n-1}$.
\EndFor
\end{algorithmic}
\end{algorithm}

After generating the offspring set $\mathcal{C}_{n-1}$, the objective values of all child individuals are evaluated in the same manner as the parent population. Environmental selection is then applied to the combined pool $\mathcal{P}_{n-1} \cup \mathcal{C}_{n-1}$, where FNS is first used to identify Pareto fronts. Complete fronts from the first one are sequentially added to the next generation, while individuals from the last partial front are selected based on their SCD, prioritizing solutions that contribute most to diversity. This process yields a new population $\mathcal{P}_n$ of size $N_{\text{pop}}$, marking the completion of a generation. The complete procedure of the modified ASDNSGA-II algorithm for HAPS placement and count joint optimization is shown in Alg.~\ref{alg:modified asdnsga ii}.

Due to the characteristics of SBX, BLX-$\alpha$, and polynomial mutation, the updated HAPS locations may occasionally fall outside the confined conical region. When this occurs, the locations are projected back to the nearest valid point within the allowable region. Additionally, since both crossover and mutation are applied probabilistically, there is a risk of generating duplicate individuals. To avoid the exclusion of any non-dominated solutions during environmental selection, uniqueness, based on both objective values and HAPS locations, is enforced among individuals during niche selection\footnote{\textit{Niche selection} is a diversity-preserving strategy in evolutionary multi-objective optimization that partitions the population into subgroups (niches), allowing the algorithm to maintain multiple distinct solution regions\cite{b39}.}, which is a component of environment selection.

\subsection{NSGA-III Algorithm}
Compared to NSGA-II, which relies on crowding distance for maintaining diversity among non-dominated solutions, NSGA-III introduces a reference-point-based selection mechanism specifically designed for many-objective optimization problems. Instead of using crowding distance, NSGA-III employs a set of well-distributed reference points in the objective space to guide the environmental selection process. 

\subsubsection{Adaptive Crossover and Polynomial Mutation} 
To ensure a fair comparative analysis, we employ the same genetic operations used in ASDNSGA-II. Specifically, an adaptive crossover switching mechanism is adopted to switch between SBX and BLX-$\alpha$, while polynomial mutation is applied to introduce additional variation. The heuristic crossover type assignment strategy and the ANND-based crowding distance calculation in the decision space, as used in ASDNSGA-II, are also incorporated into NSGA-III. In the objective space, however, the crowding distance is replaced with the perpendicular distance to the closest reference point, consistent with the core selection mechanism of NSGA-III and used to maintain solution diversity.

\subsubsection{Adaptive Perpendicular Distance Threshold}
To apply the heuristic crossover type assignment, a threshold on the perpendicular distance is required. Inspired by distance dominance relation proposed in~\cite{b29}, we compute this threshold using the $\lfloor\frac{N_{\text{pop}}}{2} \rfloor$-th smallest unique value among all perpendicular distances to reference lines. This approach promotes diversity by ensuring that roughly half of the population lies within a niche defined by the threshold. During adaptive crossover, this value is used to determine whether parent solutions are sufficiently diverse in the objective space to justify recombination. By applying this niche control mechanism, the algorithm balances convergence pressure with solution diversity, especially in high-dimensional or crowded regions of the population.

\subsection{MOPSO Algorithm}
MOPSO extends classical particle swarm optimization to multi-objective problems by guiding each particle based on its personal best and a leader chosen from an archive of non-dominated solutions. Typically, the archive preserves Pareto diversity using a hypercube-based partitioning\footnote{The \textit{hypercube-based partitioning} divides the objective space into a grid of multi-dimensional cells (hypercubes), where each cell represents a subregion or niche of the objective space.} and roulette-wheel selection to encourage exploration of underrepresented regions~\cite{b23}. However, in our problem, the discrete nature of one objective renders hypercube-based partitioning ineffective, thereby reducing the diversity of solutions in the archive and limiting the overall effectiveness of this strategy. To address this, we introduce two key modifications tailored to our problem setting.

\begin{algorithm}[htbp]
\caption{Stratified Archive Update}
\label{alg:stratified archive update}
\begin{algorithmic}[1]
\Input $\mathcal{A}_{n-1}$, $\mathcal{P}$, $N_{\text{min}}$, $N_{\text{max}}$.
\Output $\mathcal{A}_n$.
\State Combine particles $\mathcal{A}_{n-1} \cup \mathcal{P}$.
\State Compute $\mathcal{F}$ via FNS on $\mathcal{A}_{n-1} \cup \mathcal{P}$.
\Initialize $\mathcal{A}_n \gets \emptyset$.
\For{$i = N_{\text{min}} : N_{\text{max}}$}
    \State $found \gets \textbf{false}$.
    \For{$F_k$ in $\mathcal{F}$}
        \If{$found$} 
            \State \textbf{break}.
        \EndIf
        \State $C \gets$ particles in $F_k$ with HAPS count $i$.
        \If{$C \neq \emptyset$}
            \State $C_{\text{best}} \gets$ particles in $C$ with the min. PEB.
            \State $C_{\text{unique}} \gets$ remove duplicates from $C_{\text{best}}$.
            \State $\mathcal{A}_n \gets \mathcal{A}_n \cup C_{\text{unique}}$.
            \State $found \gets \textbf{true}$.
        \EndIf
    \EndFor
\EndFor
\end{algorithmic}
\end{algorithm}

\subsubsection{Stratified Archive}
To ensure compatibility during velocity updates where the HAPS count must remain consistent across the particle's personal best, its current state with the previous velocity, and the leader, we adopt a stratified archive update method, in which non-dominated solutions are grouped according to their HAPS count. 

The stratified archive update method takes four inputs: the previous archive $\mathcal{A}_{n-1}$, the new particles $\mathcal{P}$, and the minimum and maximum HAPS counts, $N_{\text{min}}$ and $N_{\text{max}}$. It returns the updated archive $\mathcal{A}_n$. The procedure begins by merging $\mathcal{A}_{n-1}$ and $\mathcal{P}$ and applying FNS to obtain Pareto fronts $\mathcal{F}$. For each HAPS count $i$ from $N_{\text{min}}$ to $N_{\text{max}}$, we traverse the fronts to find candidate solutions, known as particles, with HAPS count $i$. Once found, the subset $C$ is filtered to retain only those with the minimum PEB, stored in $C_{\text{best}}$. Duplicate particles in $C_{\text{best}}$ are removed based on identical HAPS positions, resulting in $C_{\text{unique}}$. These are then added to $\mathcal{A}_n$. The general procedure is summarized in Alg.~\ref{alg:stratified archive update}.

\subsubsection{Polynomial Mutation}
To maintain methodological consistency across all algorithms and mitigate the limited diversity resulting from the discrete nature of one objective, polynomial mutation is applied to particles after velocity updates. This mutation operator introduces additional diversity into the population and enhances the algorithm's ability to escape local optima, especially in high-dimensional or discrete decision spaces. 

\subsection{Additional Consistency Measures}
To further enhance consistency and enable a fair comparison, we 1) use the same initial population for all three algorithms; and 2) standardize the stochastic operators by using a shared per-generation crossover trigger for ASDNSGA-II and NSGA-III and a shared per-generation mutation trigger for all algorithms.

\section{Simulation Setup}

\begin{figure*}[t]
\centering
\begin{subfigure}[t]{0.24\textwidth}
    \centering
    \includegraphics[width=\linewidth,height=4cm, keepaspectratio]{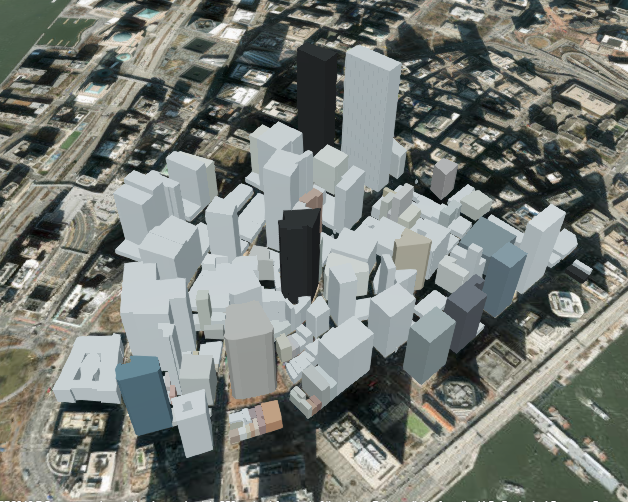}
    \caption{Wall Street.}
    \label{fig:city model_wallstreet}
\end{subfigure}
\hfill
\begin{subfigure}[t]{0.28\textwidth}
    \centering
    \includegraphics[width=\linewidth,height=4cm, keepaspectratio]{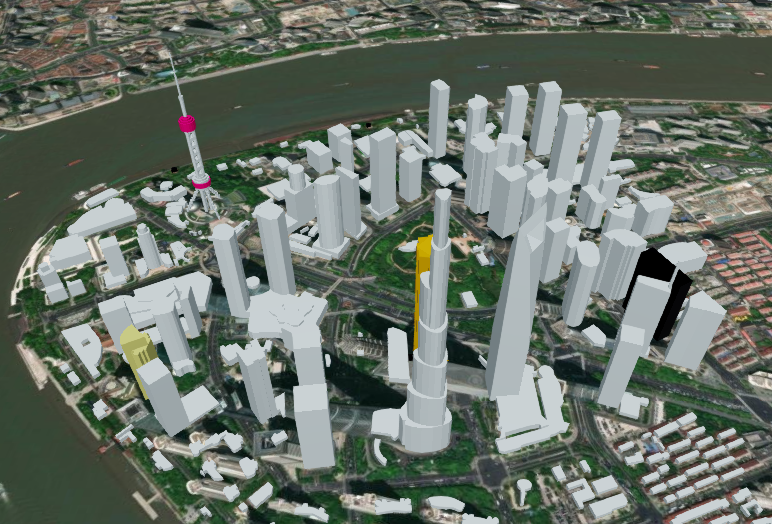}
    \caption{Lujiazui.}
    \label{fig:city model_lujiazui}
\end{subfigure}
\hfill
\begin{subfigure}[t]{0.37\textwidth}
    \centering
    \includegraphics[width=\linewidth,height=4cm, keepaspectratio]{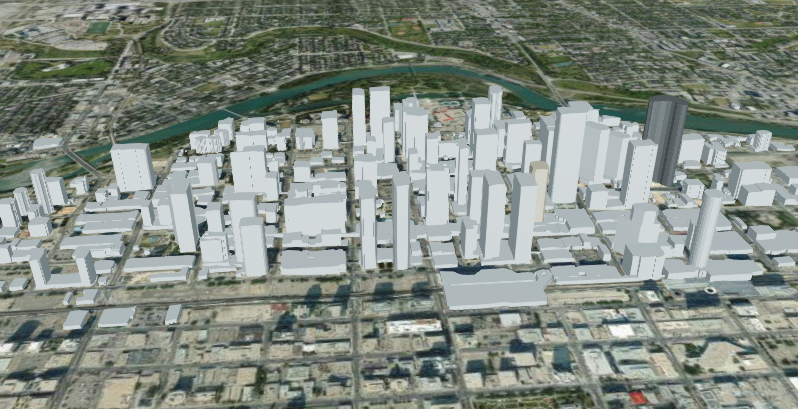}
    \caption{Downtown Calgary.}
    \label{fig:city model_calgary}
\end{subfigure}
\caption{3D city models of three urban regions: Wall Street (New York City), Lujiazui (Shanghai), and downtown Calgary.}
\label{fig:city model}
\end{figure*}

Without loss of generality and to ensure the adaptability of the proposed framework, three representative urban regions are considered in this study: the Wall Street area in New York City, Lujiazui in Shanghai, and downtown Calgary. While Lujiazui exhibits a similar level of GPS coverage to Wall Street, its urban layout differs significantly; in particular, it lacks a dense concentration of skyscrapers in the central area. In contrast, downtown Calgary features wider streets and shorter buildings, but experiences relatively poorer GPS coverage than the other two regions, primarily due to its higher latitude. The corresponding 3D city models are shown in Fig.~\ref{fig:city model}.

\begin{figure*}[t]
\centering
\begin{subfigure}[t]{0.32\textwidth}
    \centering
    \includegraphics[width=\linewidth]{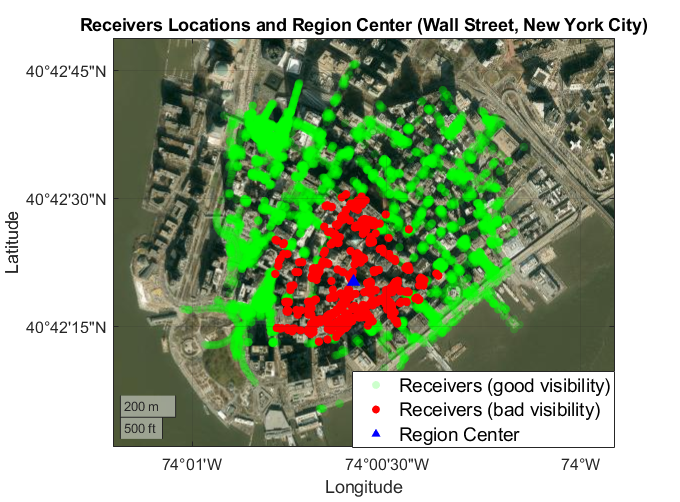}
    \caption{Wall Street.}
    \label{fig:rec_wallstreet}
\end{subfigure}
\hfill
\begin{subfigure}[t]{0.32\textwidth}
    \centering
    \includegraphics[width=\linewidth]{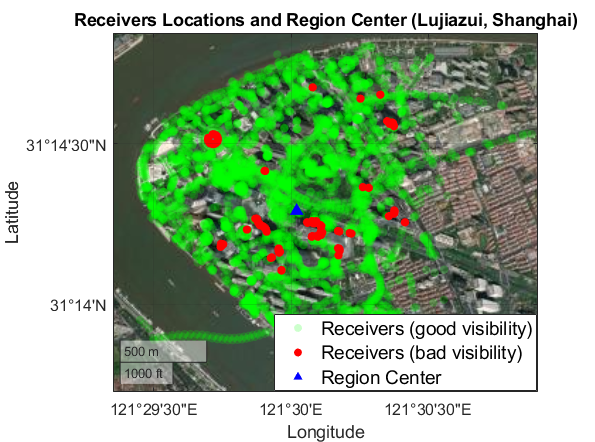}
    \caption{Lujiazui.}
    \label{fig:rec_lujiazui}
\end{subfigure}
\hfill
\begin{subfigure}[t]{0.32\textwidth}
    \centering
    \includegraphics[width=\linewidth]{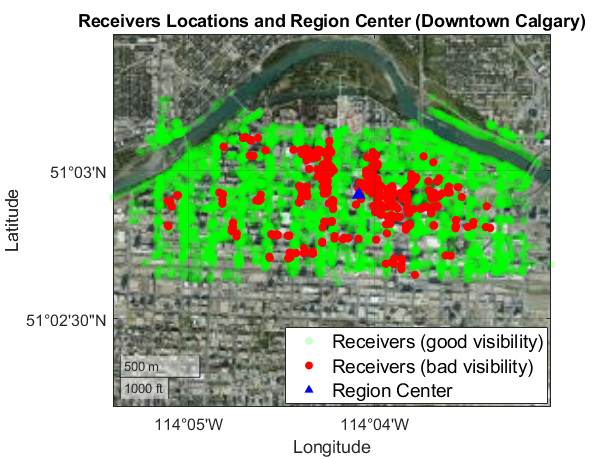}
    \caption{Downtown Calgary.}
    \label{fig:rec_calgary}
\end{subfigure}
\caption{Spatial distribution of receiver locations in Wall Street (New York City), Lujiazui (Shanghai), and downtown Calgary.}
\label{fig:receiver locations}
\end{figure*}

Based on the OpenStreetMap (OSM) data, all street-level nodes in the cropped urban areas, which are the nodes associated with elements tagged as highway, are used as candidate receiver locations. Given the large number of street nodes and the complexity of the building geometry, repeatedly performing LOS/NLOS determination via ray tracing for every transmitter–receiver pair within the main loop would be computationally prohibitive. To mitigate this cost, we precompute a skyline mask for each candidate at a $1^\circ$ azimuth–elevation resolution using the Intel Embree ray-tracing engine. To further reduce computation, the analysis is restricted to severely blocked regions by retaining only those candidates whose mean skyline elevation exceeds 50°, indicating limited sky visibility. This results in $N_{\text{r}} = 932$ bad-visibility receivers in Wall Street, $N_{\text{r}} = 135$ in Lujiazui, and $N_{\text{r}} = 489$ in downtown Calgary, which define the ROIs used in the subsequent simulations. The spatial distributions of both good- and bad-visibility receivers and their corresponding region centers are shown in Fig.~\ref{fig:receiver locations}.

% \begin{figure}[!t]
% \centering
% \includegraphics[width=0.9\columnwidth]{images/receiver locations, wallstreet.png}
% \caption{Receiver locations in the vicinity of Wall Street, New York City, USA.}
% \label{fig:receiver locations}
% \end{figure}

The satellite scenario is created using MATLAB's Satellite Communications Toolbox, supplemented by a two-line element (TLE) file. In this study, the WGS-84 Earth model is implemented consistently across all simulation modules, and a commonly used elevation mask of 10$^\circ$ is applied for both satellites and HAPS~\cite{b2}. Because satellite geometry changes over a sidereal day, evaluating all possible snapshots would be computationally prohibitive. Accordingly, we first compute the average 3D PEB across the ROI for the satellite-only case at 15-minute intervals, yielding 97 snapshots. From these, only the $N_{\text{t}} = 10$ snapshots with the highest average PEB are retained. This snapshot-selection strategy substantially reduces the computational burden while concentrating the analysis on the most challenging geometric conditions. The full metaheuristic framework, which simultaneously evaluates the three considered metaheuristic algorithms, requires approximately 1 hour for Wall Street, 15 minutes for Lujiazui, and 35 minutes for downtown Calgary on a PC equipped with 6 CPU cores and 64 GB of RAM.

\begin{table*}[t]
\caption{Simulation parameters for ASDNSGA-II, NSGA-III, and MOPSO.}
\begin{center}
\renewcommand{\arraystretch}{1.3}
\setlength{\tabcolsep}{6pt}
\begin{tabular}{>{\centering\arraybackslash}m{0.15\linewidth}|
                >{\centering\arraybackslash}m{0.16\linewidth}|
                >{\centering\arraybackslash}m{0.12\linewidth}||
                >{\centering\arraybackslash}m{0.16\linewidth}|
                >{\centering\arraybackslash}m{0.12\linewidth}}
\Xhline{1pt}
\textbf{Category} & \textbf{Parameter} & \textbf{Value} & \textbf{Parameter} & \textbf{Value} \\
\Xhline{1pt}
\multirow{2}{*}{General}
  & $\theta_{\text{min}}$ & 10$^\circ$ & $\tau$ & 18 m \\
  & $N_{\text{g}}$ & 200 & -- & -- \\
\hline
\multirow{5}{*}{\shortstack{ASDNSGA-II, \\ NSGA-III}}
  & $p_{\text{c}}$ & 0.9 & $p_{\text{m}}$ & 0.1 \\
  & $\eta_{\text{c}}$ & 20 & $\eta_{\text{m}}$ & 20 \\
  & $N_{\text{pop}}$ & 100 & $N_{\text{min}}$ & 1 \\
  & $N_{\text{max}}$ & 8 & $d_{\text{dec}}^{\text{th}}$ & [0.3, 0.5, 0.7] \\
  & $d_{\text{obj, ASDNSGA-II}}^{\text{th}}$ & [0.3, 0.5, 0.7] & -- & -- \\
\hline
\multirow{2}{*}{MOPSO}
  & $w$ & 0.5 & $c_1$ & 1.5 \\
  & $c_2$ & 2.0 & -- & -- \\
\Xhline{1pt}
\end{tabular}
\label{tab:simulation parameters}
\end{center}
\end{table*}

Ensuring a fair comparison across all algorithms requires carefully selected parameters. For ASDNSGA-II and NSGA-III, the crossover probability $p_{\text{c}}$, mutation probability $p_{\text{m}}$, and the auxiliary parameters for SBX and polynomial mutation, $\eta_{\text{c}}$ and $\eta_{\text{m}}$, respectively, are configured following the settings reported in~\cite{b27}. For MOPSO, the inertia weight $w$ is commonly chosen between 0.4 and 0.9, while the cognitive acceleration coefficient $c_1$ and the social acceleration coefficient $c_2$ are typically set between 1.5 and 2~\cite{b37}. In this work, we select $w=0.5$, $c_1=1.5$, and $c_2=2$ to strike a balance between exploration and exploitation. The minimum and maximum HAPS count are set to 1 and 8, respectively. To evaluate the impact of heuristic thresholds on solution quality, decision-space crowding-distance thresholds of 0.3, 0.5, and 0.7 are adopted for both ASDNSGA-II and NSGA-III. In the objective space, ASDNSGA-II employs the same set of thresholds, whereas NSGA-III uses the adaptive thresholding method described in Section~\Romannum{3}. Given the urban environment under consideration, the PEB threshold $\tau$ is set to 18 m. A comprehensive summary of all simulation parameters is provided in Table~\ref{tab:simulation parameters}.

To provide a simple and interpretable benchmark for comparison against the developed metaheuristic solutions, we implement a fingerprinting-based baseline, namely the Greedy-Add algorithm. Greedy-Add has been widely applied in sensor- and anchor-placement problems and remains a valid baseline even when effective noise levels vary across candidate locations~\cite{b40,b41}. This is partly due to the diminishing-returns behavior common in ranging-based systems, where the incremental benefit of adding additional anchors decreases as the geometry becomes saturated. Greedy-Add evaluates the marginal reduction in the average 3D PEB achieved by adding each candidate HAPS and iteratively selects the best location until no further improvement is observed or the maximum number of HAPS is reached.

Balancing geometric fidelity and computational efficiency requires discretizing the candidate HAPS space into a curvature-aware conical grid above the region center with well-calibrated angular and altitude resolutions. In this work, a $0.1^\circ$ azimuth-elevation resolution and 1 km altitude steps are used, providing a dense enough sampling of feasible HAPS geometries to capture meaningful PEB variations, yet remain computationally manageable for repeated evaluation within Greedy-Add and the proposed metaheuristics. The grid also enforces the defined constraints, including the $10^\circ$ elevation mask and an altitude range of 18-22 km. This deterministic and parameter-free baseline provides a transparent reference against which the performance gains of the developed metaheuristic solutions can be assessed.

Despite the chosen granularity, the search space remains extremely large, containing a total of 14.4 million candidate solutions. To accelerate computation, the Greedy-Add algorithm is executed on a high-performance computing (HPC) cluster with 192 CPU cores and 750 GB of RAM.

\section{Simulation Results}
This section provides a comprehensive analysis of algorithmic performance, system performance, GDOP behavior, and runtime bottlenecks, focusing on the Wall Street scenario, followed by a generalization analysis of two other city models.

\begin{figure*}[t]
\centering
\begin{subfigure}[t]{0.24\textwidth}
    \centering
    \includegraphics[width=\linewidth]{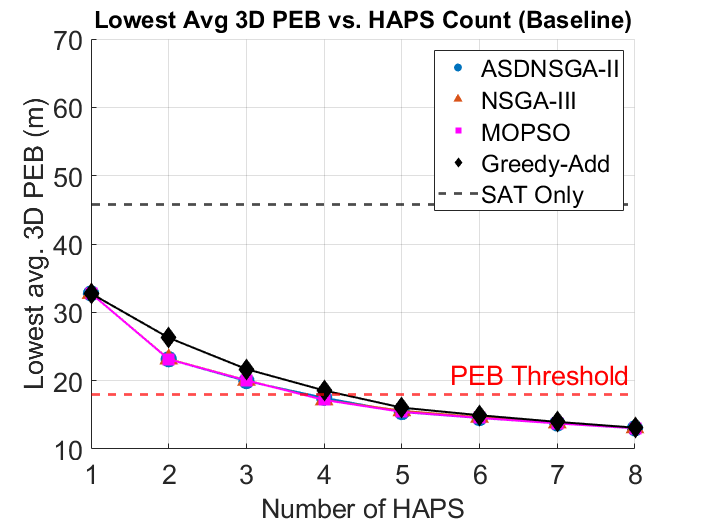}
    \caption{Baseline.}
    \label{fig:baseline}
\end{subfigure}
\hfill
\begin{subfigure}[t]{0.24\textwidth}
    \centering
    \includegraphics[width=\linewidth]{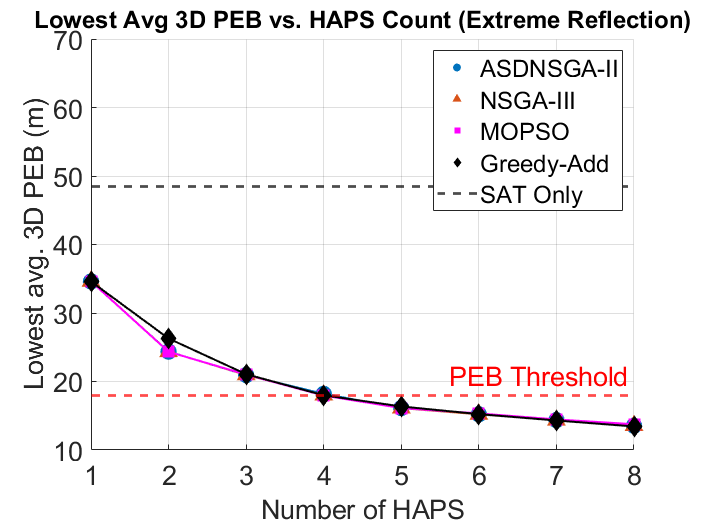}
    \caption{Extreme Reflection.}
    \label{fig:extreme_ref}
\end{subfigure}
\hfill
\begin{subfigure}[t]{0.24\textwidth}
    \centering
    \includegraphics[width=\linewidth]{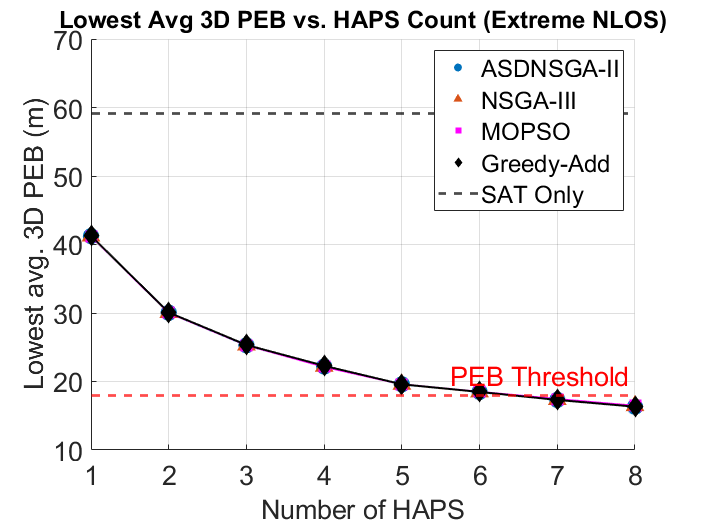}
    \caption{Extreme NLOS.}
    \label{fig:extreme_nlos}
\end{subfigure}
\hfill
\begin{subfigure}[t]{0.24\textwidth}
    \centering
    \includegraphics[width=\linewidth]{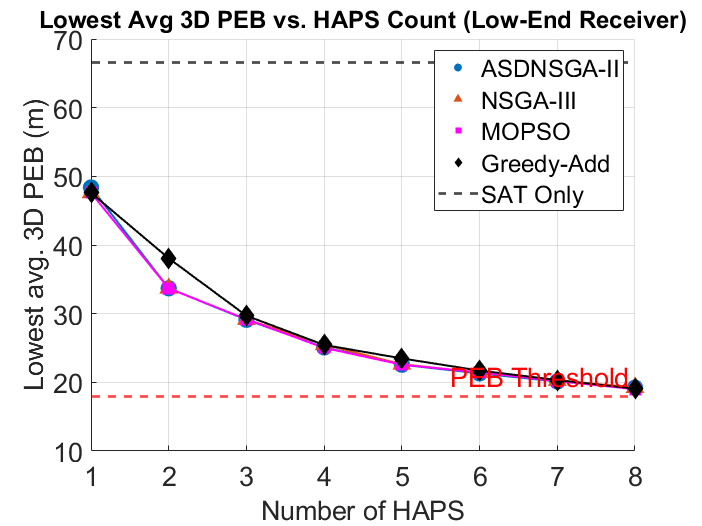}
    \caption{Low-End Receiver.}
    \label{fig:low_end_rec}
\end{subfigure}
\caption{Final-generation Pareto fronts of average 3D PEB versus HAPS count for all algorithms under Baseline, Extreme NLOS, Extreme Reflection, and Low-End Receiver scenarios.}
\label{fig:lowest_avg_peb_vs_haps_count_final_gmm}
\end{figure*}

\subsection{Algorithmic Performance Analysis}
The algorithmic performance is evaluated in terms of robustness and convergence behavior. For robustness, we first examine performance across diverse urban environments characterized by different GMM parameters. We then analyze robustness with respect to variations in the crowding distance thresholds. For the subsequent convergence analysis, we fix both $d_{\text{dec}}^{\text{th}}$ and $d_{\text{obj}}^{\text{th}}$ at 0.5 and adopt the GMM parameter set corresponding to the baseline scenario.

\subsubsection{Robustness Analysis Under Varying GMM Parameters}
The robustness of the proposed metaheuristic framework under diverse urban scenarios is evaluated using four sets of GMM parameters. Fig.~\ref{fig:lowest_avg_peb_vs_haps_count_final_gmm} presents the final-generation Pareto fronts of average 3D PEB versus HAPS count for all algorithms under the four considered scenarios: Baseline, Extreme NLOS, Extreme Reflection, and Low-End Receiver. The results show that the developed metaheuristic solutions (ASDNSGA-II, NSGA-III, and MOPSO) achieve consistently competitive performance compared to the Greedy-Add baseline, particularly in the low-to-moderate HAPS regime (e.g., 2-6 HAPS). In this range, the metaheuristics generally attain lower average 3D PEB, although the margin varies across scenarios. As the HAPS count increases, the performance gap diminishes and all methods converge to nearly identical solutions, indicating that the benefit of global search is most pronounced under limited deployment budgets, where the placement problem is more constrained.

Environmental conditions are observed to strongly influence both the achievable PEB and the minimum number of HAPS required to meet a target PEB threshold. This is because harsher conditions, such as increased NLOS blockage, strong multipath reflections, or degraded receiver quality, effectively reduce measurement reliability and geometric diversity. In such cases, additional HAPS are needed to compensate for poor satellite visibility, mitigate unfavorable geometry, and provide more robust ranging links, thereby increasing the infrastructure requirement.

Despite these variations in channel and receiver conditions, the proposed metaheuristic framework exhibits strong robustness across all considered urban scenarios, corresponding to different GMM parameter settings. In each case, the algorithms consistently converge to similar high-quality Pareto fronts and maintain their relative performance advantage over the greedy approach in low-to-moderate HAPS regimes. This indicates that the framework effectively adapts to changes in propagation characteristics and measurement uncertainty, making it suitable for a wide range of dense urban environments.

\begin{figure*}[t]
\centering
\begin{subfigure}[t]{0.48\textwidth}
    \centering
    \includegraphics[width=\linewidth]{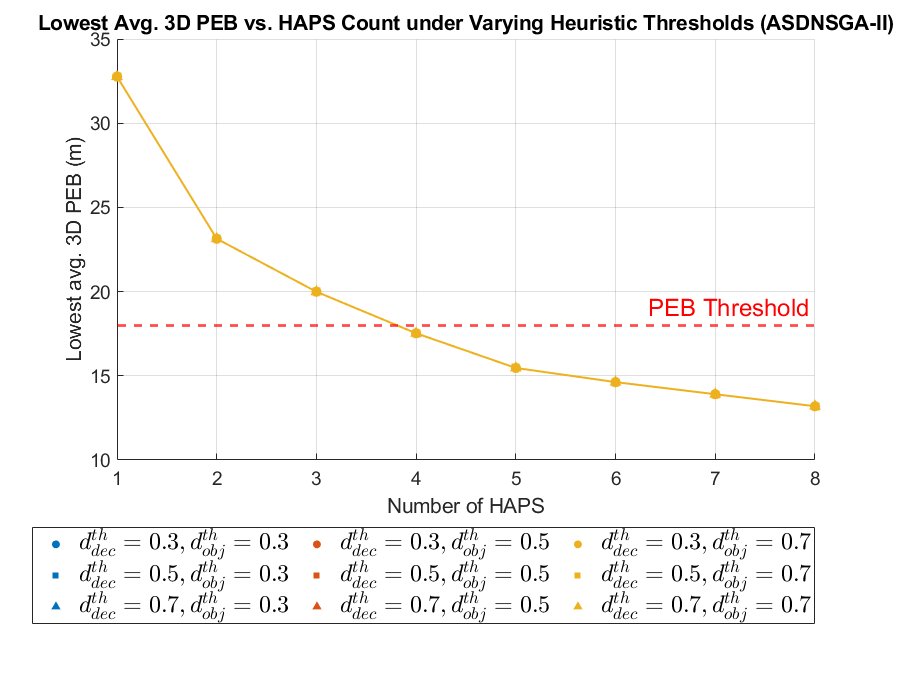}
    \caption{ASDNSGA-II.}
    \label{fig:heu_th_asdnsga2}
\end{subfigure}
\hfill
\begin{subfigure}[t]{0.48\textwidth}
    \centering
    \includegraphics[width=\linewidth]{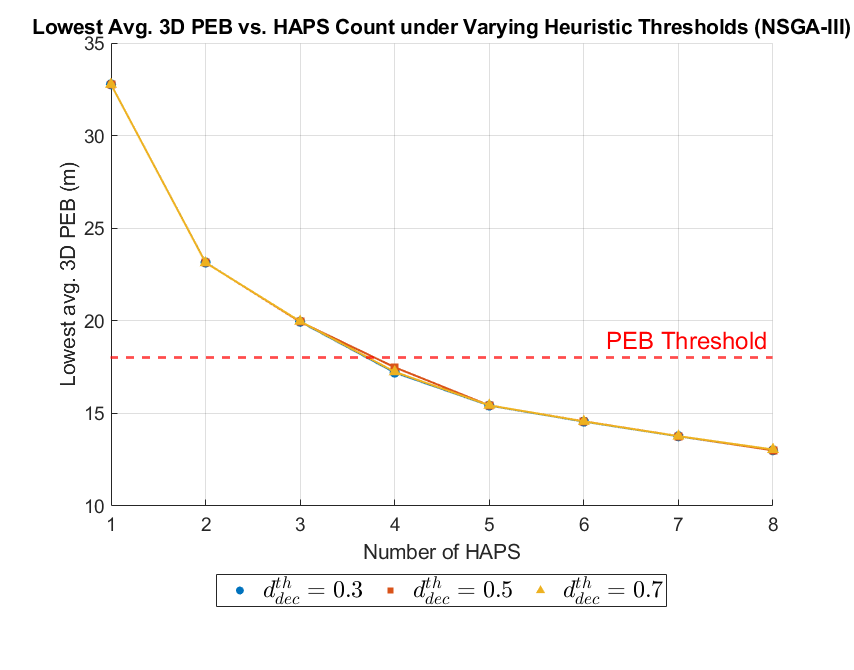}
    \caption{NSGA-III.}
    \label{fig:heu_th_nsga3}
\end{subfigure}
\caption{Final-generation Pareto fronts of average 3D PEB versus HAPS count for ASDNSGA-II and NSGA-III under under varying crowding distance thresholds.}
\label{fig:lowest_avg_peb_vs_haps_count_final_heu_th}
\end{figure*}

\subsubsection{Robustness Analysis Under Varying Crowding Distance Thresholds}
As ASDNSGA-II and NSGA-II employ an adaptive crossover switching mechanism, governed by the crowding distance thresholds $d_{\text{dec}}^{\text{th}}$ and $d_{\text{obj}}^{\text{th}}$, to balance exploration and exploitation, their robustness is further evaluated under varying threshold settings. Fig.~\ref{fig:lowest_avg_peb_vs_haps_count_final_heu_th} presents the final-generation Pareto fronts of average 3D PEB versus HAPS count under varying crowding distance thresholds for both GAs. Note that for NSGA-III, the crowding distance in the objective space $d_{\text{obj}}^{\text{th}}$ is replaced by the perpendicular distance to the closest reference point, for which an adaptive perpendicular distance threshold is employed; therefore, only variations in $d_{\text{dec}}^{\text{th}}$ are considered. Based on Fig.~\ref{fig:lowest_avg_peb_vs_haps_count_final_heu_th}, we can see that both algorithms exhibit nearly identical performance across all tested combinations of $d_{\text{dec}}^{\text{th}}$ and $d_{\text{obj}}^{\text{th}}$, indicating that the optimized solution quality is largely insensitive to the specific choice of threshold values within the considered range.

\begin{figure*}[t]
\centering
\begin{subfigure}[t]{0.48\textwidth}
    \centering
    \includegraphics[width=\linewidth]{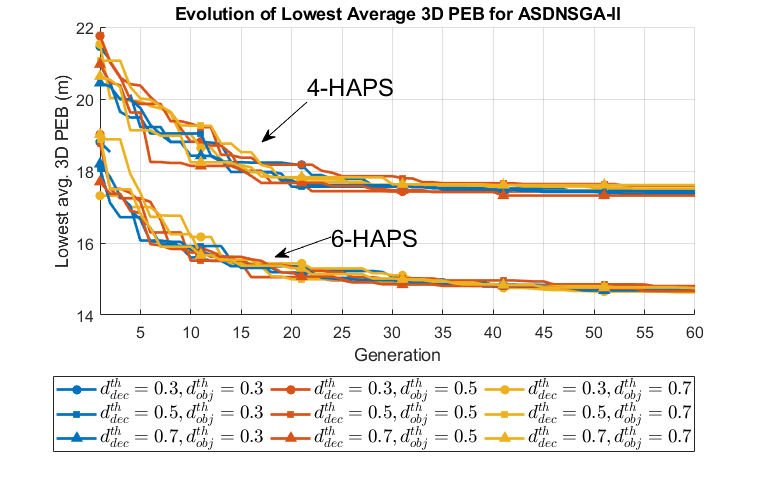}
    \caption{ASDNSGA-II.}
    \label{fig:evolution_asdnsga2_heu_th}
\end{subfigure}
\hfill
\begin{subfigure}[t]{0.48\textwidth}
    \centering
    \includegraphics[width=\linewidth]{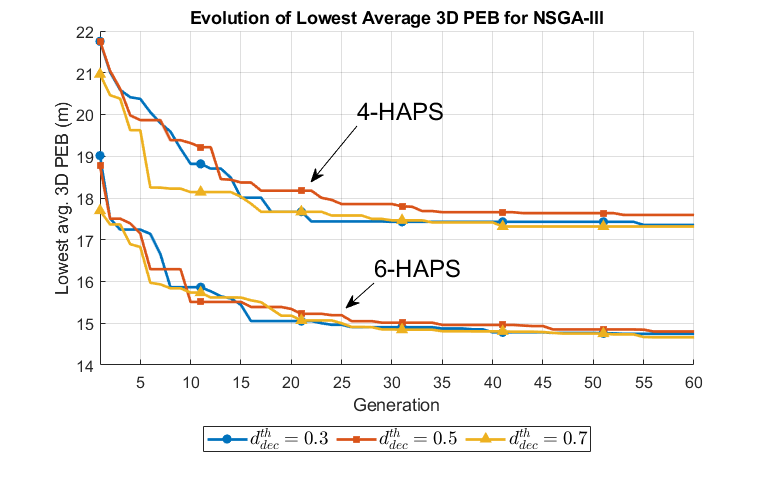}
    \caption{NSGA-III.}
    \label{fig:evolution_nsga3_heu_th}
\end{subfigure}
\caption{Evolution of the lowest average 3D PEB over generations under different heuristic thresholds for ASDNSGA-II and NSGA-III.}
\label{fig:evolution_lowest_avg_peb_vs_haps_count_heu_th}
\end{figure*}

Since the crowding distance thresholds influence the balance between exploration and exploitation, we also evaluate their impact on convergence speed. Fig.~\ref{fig:evolution_lowest_avg_peb_vs_haps_count_heu_th} shows the evolution of the lowest average 3D PEB over generations under different heuristic thresholds for both algorithms, using the 4-HAPS and 6-HAPS cases as representative examples. The results show that both algorithms converge at around generation 50, with minor variations in the convergence trajectories primarily attributable to stochastic effects. This further suggests that the developed framework is largely insensitive to the choice of crowding distance thresholds. This robustness can be attributed to the combined effect of multiple operators, including improved binary tournament selection, polynomial mutation, and environmental selection, which collectively maintain a good balance between convergence and diversity.

\subsubsection{Convergence Behavior Analysis}
To evaluate the convergence behavior of the considered algorithms, we track the best configuration of each algorithm in each generation. The selection of the best configuration is based on the optimization objectives defined in Eq.\ref{eq:opt1} and Eq.\ref{eq:opt2}. For both ASDNSGA-II and NSGA-III, elite selection is performed on the first Pareto front, which contains the non-dominated solutions representing the best trade-offs between the conflicting objectives. Within this front, solutions with an average 3D PEB below the predefined threshold are first identified. Among these, the one with the fewest HAPS is selected as the elite. For MOPSO, which employ an external archive of non-dominated solutions, the archive serves the role of the first front. Accordingly, the elite solution is selected from the archive using the same two-stage criterion.

\begin{figure}[!t]
\centering
\includegraphics[width=0.9\columnwidth]{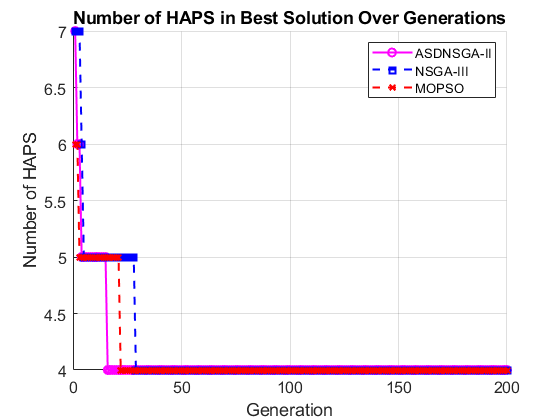}
\caption{HAPS count in best solution over generations for ASDNSGA-II, NSGA-III, and MOPSO.}
\label{fig:number of haps in best solution over generations (ASDNSGA-II, NSGA-III, MOPSO)}
\end{figure}

Based on the specified simulation settings, we evaluate the ability of ASDNSGA-II, NSGA-III, and MOPSO in identifying optimal solutions for both HAPS placement and count. Fig.~\ref{fig:number of haps in best solution over generations (ASDNSGA-II, NSGA-III, MOPSO)} illustrates the evolution of HAPS count in the best solution across generations. As shown in the figure, all three metaheuristic algorithms exhibit a similar convergence trend: the number of HAPS is progressively reduced as the search proceeds. Ultimately, all three algorithms converge to a 4-HAPS configuration, indicating that at least four HAPS are needed to augment the satellite system and achieve an average 3D PEB below the predefined threshold.

\begin{figure}[t]
    \centering

    \begin{subfigure}{0.48\linewidth}
        \centering
        \includegraphics[width=\linewidth]{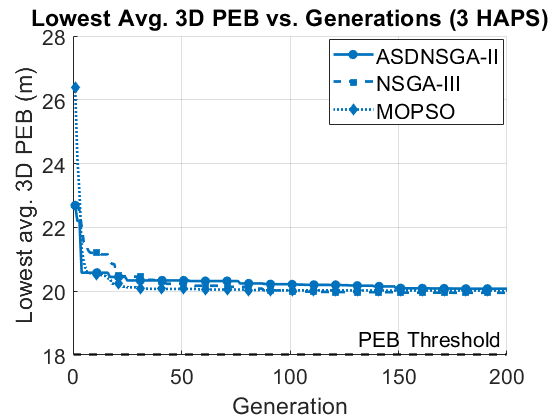}
        \caption{3-HAPS.}
    \end{subfigure}
    \hfill
    \begin{subfigure}{0.48\linewidth}
        \centering
        \includegraphics[width=\linewidth]{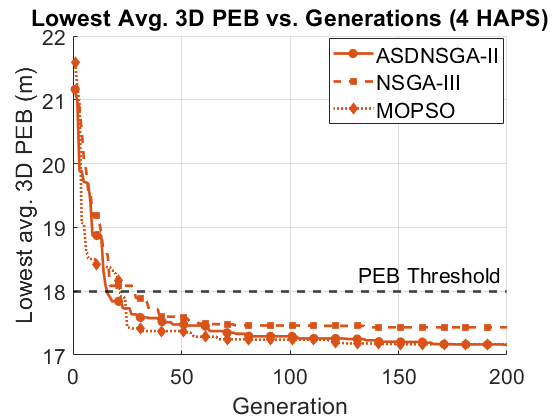}
        \caption{4-HAPS.}
    \end{subfigure}

    \vspace{0.25cm}

    \begin{subfigure}{0.48\linewidth}
        \centering
        \includegraphics[width=\linewidth]{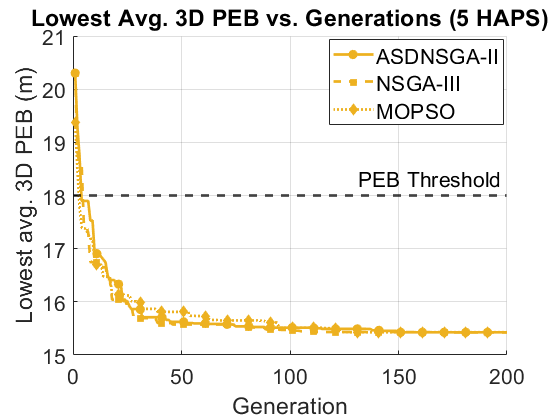}
        \caption{5-HAPS.}
    \end{subfigure}
    \hfill
    \begin{subfigure}{0.48\linewidth}
        \centering
        \includegraphics[width=\linewidth]{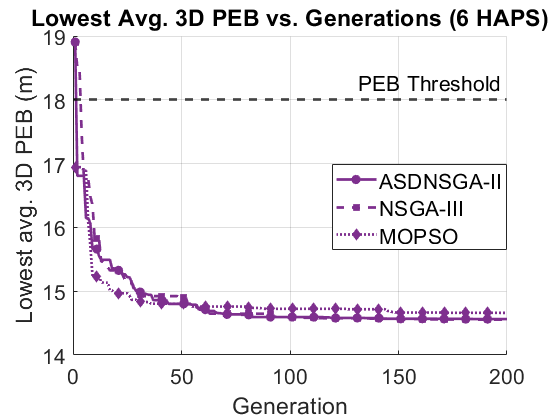}
        \caption{6-HAPS.}
    \end{subfigure}

    \caption{Evolution of the lowest average 3D PEB over generations for different HAPS counts using ASDNSGA-II, NSGA-III, and MOPSO.}
    \label{fig:lowest avg 3d peb vs gen}
\end{figure}

To assess the convergence behavior of the three metaheuristic algorithms, Fig.~\ref{fig:lowest avg 3d peb vs gen} illustrates the evolution of the lowest average 3D PEB achieved over generations for selected HAPS counts, specifically 3, 4, 5, and 6. The results show that all three algorithms converge rapidly, with most of the reduction in the lowest average 3D PEB occurring within the first 50 generations. 

% \begin{figure}[!t]
% \centering
% \includegraphics[width=0.9\columnwidth]{images/lowest avg 3d peb vs haps count_final.png}
% \caption{Lowest average 3D PEB vs HAPS count in the final generation for ASDNSGA-II, NSGA-III, and MOPSO (solid lines: lowest average 3D PEB; dashed line: average 3D PEB).}
% \label{fig:lowest avg 3d peb vs haps count_final}
% \end{figure}

% Fig.~\ref{fig:lowest avg 3d peb vs haps count_final} shows the lowest average 3D PEB obtained for each HAPS count in the final generation across the three metaheuristic algorithms, together with the Greedy-Add baseline. The satellite-only case is also included for reference. As seen in the figure, the three metaheuristic algorithms deliver nearly identical performance for all HAPS counts, indicating that—despite their different search principles—they are equally effective once the methodological consistency measures described earlier are applied.

\subsection{System Performance Analysis}
\begin{figure*}[t]
\centering
\begin{subfigure}[t]{0.32\textwidth}
    \centering
    \includegraphics[width=\linewidth]{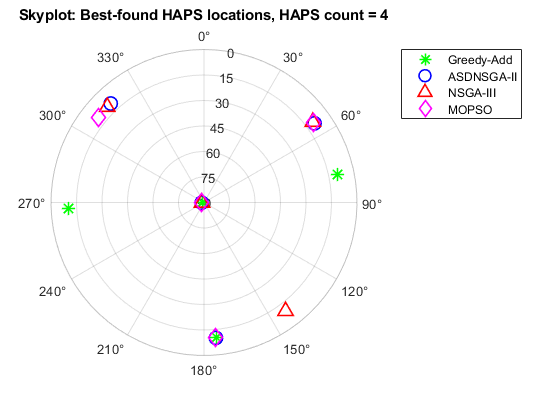}
    \caption{4-HAPS configuration.}
    \label{fig:skyplot (4-haps)}
\end{subfigure}
\hfill
\begin{subfigure}[t]{0.32\textwidth}
    \centering
    \includegraphics[width=\linewidth]{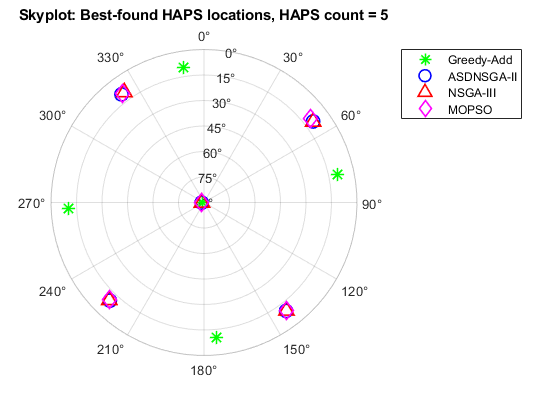}
    \caption{5-HAPS configuration.}
    \label{fig:skyplot (5-haps)}
\end{subfigure}
\hfill
\begin{subfigure}[t]{0.32\textwidth}
    \centering
    \includegraphics[width=\linewidth]{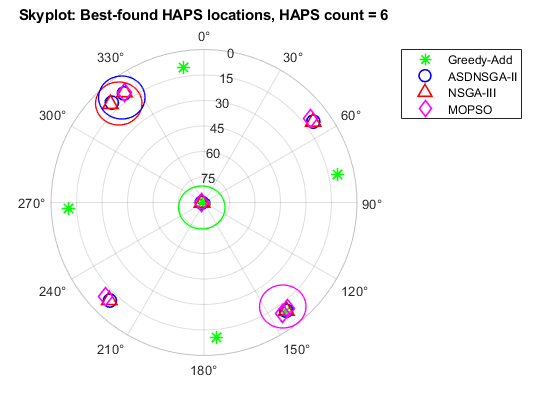}
    \caption{6-HAPS configuration.}
    \label{fig:skyplot (6-haps)}
\end{subfigure}
\caption{Skyplot illustrating the optimal HAPS locations for 4-, 5-, and 6-HAPS configurations. For the 6-HAPS case, overlap regions are highlighted with colored circles: \textcolor{blue}{blue} (ASDNSGA-II), 
\textcolor{blue}{red} (NSGA-III), 
\textcolor{magenta}{magenta} (MOPSO), 
\textcolor{green}{green} (Greedy-Add).}
\label{fig:skyplot}
\end{figure*}

With the optimal HAPS configurations identified for each HAPS count, we now evaluate the system-level performance and robustness of the positioning system over a full sidereal day. Since the PEB threshold is satisfied with four HAPS and additional HAPS yield only diminishing returns, Fig.~\ref{fig:skyplot} presents the skyplots of the best HAPS configurations for 4, 5, and 6 HAPS generated by ASDNSGA-II, NSGA-III, MOPSO, and Greedy-Add. Across all algorithms, a HAPS is consistently positioned directly above the ROI. This placement is expected, as a nadir-pointing HAPS maintains a high elevation angle to most receivers, thereby ensuring LOS visibility with a large subset of them. The remaining HAPS, except in the 4-HAPS case of Greedy-Add, are generally well distributed azimuthally around the region. Such wide angular separation contributes to strong geometric diversity: the central HAPS primarily improves vertical dilution of precision (VDOP), while the lower-elevation HAPS located around the perimeter more directly reduce horizontal dilution of precision (HDOP). Notably, none of the metaheuristic algorithms place a HAPS in the azimuth sector that is underrepresented by GPS satellites. This may be due to the presence of tall buildings in the northern part of the cropped city model, which restrict LOS opportunities and reduce the potential geometric benefit of placing a HAPS in that direction.

Some variability among the metaheuristic solutions is also expected. For example, in the 4-HAPS case, NSGA-III identifies one HAPS position that differs from those found by ASDNSGA-II and MOPSO, yet all three yield nearly identical PEB performance. This illustrates an important point: due to the discretized search grid and the complex interplay among LOS availability, geometric dilution, and urban obstructions, multiple distinct configurations can achieve comparable accuracy.

Finally, in the 6-HAPS case, all algorithms begin to exhibit placement overlaps, where at least one pair of HAPS occupy nearby azimuth–elevation regions. This convergence toward redundant geometry explains the observed diminishing returns at higher HAPS counts and the ability of Greedy-Add to close the performance gap with the metaheuristic algorithms. From a broader system-design perspective, such overlaps are also undesirable for communication-focused HAPS deployments, as they lead to inefficient spatial coverage and reduced geometric diversity. This further supports the conclusion that adding more HAPS beyond five provides limited additional benefit in both positioning performance and practical deployment efficiency.

% \begin{table}[t]
% \caption{Reduction in average 3D PEB Relative to Satellite-Only for Varying HAPS Counts.}
% \begin{center}
% \renewcommand{\arraystretch}{1.3} % Add vertical padding
% \setlength{\tabcolsep}{6pt} % Control horizontal padding
% \begin{tabular}{>{\centering\arraybackslash}m{0.1\linewidth}|>{\centering\arraybackslash}m{0.2\linewidth}|>{\centering\arraybackslash}m{0.2\linewidth}|>{\centering\arraybackslash}m{0.2\linewidth}}
% \Xhline{1pt}
% \textbf{HAPS Count} & ASDNSGA-II & NSGA-III & MOPSO\\
% \Xhline{1pt}
% 1 & 22.48\% & 22.77\% & 22.48\%\\
% \hline
% 2 & 38.87\% & 37.98\% & 34.59\%\\
% \hline
% 3 & 51.58\% & 50.96\% & 45.91\%\\
% \hline
% 4 & 60.67\% & 60.42\% & 51.83\%\\
% \hline
% 5 & 65.03\% & 64.52\% & 59.10\%\\
% \hline
% 6 & 67.37\% & 67.19\% & 60.97\%\\
% \hline
% 7 & 69.51\% & 69.30\% & 60.82\%\\
% \Xhline{1pt}
% \end{tabular}
% \label{tab:performance gain}
% \end{center}
% \end{table}

\begin{table*}[t]
\caption{Performance Evaluation of Average 3D PEB for Different Algorithms Over One Sidereal Day.}
\begin{center}
\renewcommand{\arraystretch}{1.3}
\setlength{\tabcolsep}{6pt}

\begin{tabular}{%
    >{\centering\arraybackslash}m{0.14\linewidth}|
    >{\centering\arraybackslash}m{0.14\linewidth}|
    *{8}{>{\centering\arraybackslash}m{0.06\linewidth}}}

\Xhline{1pt}
\textbf{Metric} & \textbf{Algorithm} 
    & \textbf{1} & \textbf{2} & \textbf{3} & \textbf{4}
    & \textbf{5} & \textbf{6} & \textbf{7} & \textbf{8} \\
\Xhline{1pt}

\multirow{4}{*}{Mean [m]}
  & Greedy-Add   & 27.186 & 22.691 & 19.900 & 17.254 & 15.228 & 14.339 & 13.459 & 12.661 \\
  & ASDNSGA-II   & 27.237 & 20.960 & 18.672 & 16.180 & 14.746 & 13.951 & 13.280 & 12.619 \\
  & NSGA-III     & 27.243 & 20.932 & 18.341 & 16.439 & 14.738 & 13.942 & 13.263 & 12.561 \\
  & MOPSO        & 27.244 & 20.925 & 18.627 & 16.182 & 14.745 & 14.113 & 13.385 & 12.682 \\
\hline

\multirow{4}{*}{STD [m]}
  & Greedy-Add   & 3.412 & 1.773 & 1.224 & 0.897 & 0.421 & 0.389 & 0.295 & 0.270 \\
  & ASDNSGA-II   & 3.475 & 1.261 & 0.801 & 0.515 & 0.372 & 0.322 & 0.276 & 0.220 \\
  & NSGA-III     & 3.486 & 1.250 & 0.983 & 0.630 & 0.374 & 0.323 & 0.275 & 0.244 \\
  & MOPSO        & 3.488 & 1.250 & 0.792 & 0.512 & 0.370 & 0.362 & 0.273 & 0.227 \\
\hline

\multirow{4}{*}{RMS [m]}
  & Greedy-Add   & 27.400 & 22.760 & 19.938 & 17.277 & 15.234 & 14.344 & 13.462 & 12.664 \\
  & ASDNSGA-II   & 27.458 & 20.998 & 18.689 & 16.189 & 14.751 & 13.955 & 13.283 & 12.621 \\
  & NSGA-III     & 27.465 & 20.970 & 18.368 & 16.451 & 14.743 & 13.946 & 13.265 & 12.563 \\
  & MOPSO        & 27.467 & 20.963 & 18.644 & 16.190 & 14.750 & 14.118 & 13.388 & 12.684 \\
\hline

\multirow{4}{*}{CV}
  & Greedy-Add   & 0.125 & 0.078 & 0.062 & 0.052 & 0.028 & 0.027 & 0.022 & 0.021 \\
  & ASDNSGA-II   & 0.128 & 0.060 & 0.043 & 0.032 & 0.025 & 0.023 & 0.021 & 0.017 \\
  & NSGA-III     & 0.128 & 0.060 & 0.054 & 0.038 & 0.025 & 0.023 & 0.021 & 0.019 \\
  & MOPSO        & 0.128 & 0.060 & 0.043 & 0.032 & 0.025 & 0.026 & 0.020 & 0.018 \\
\Xhline{1pt}

\end{tabular}

\label{tab:peb_performance}
\end{center}
\end{table*}

Next, we evaluate the accuracy and robustness of the overall system, combining both the available satellites and the optimal HAPS configurations, across 97 snapshots sampled at 15-minute intervals over one full sidereal day. For each considered HAPS count, we compute the mean, standard deviation (STD), root mean square (RMS), and coefficient of variation (CV) of the average 3D PEB. The resulting statistics across all HAPS counts are summarized in Table~\ref{tab:peb_performance}. For reference, the satellite-only baseline yields a mean of 32.2059 m, an STD of 6.0677 m, an RMS of 32.7667 m, and a CV of 0.1884 for the average 3D PEB computed across snapshots.

\begin{figure}[t]
    \centering

    \begin{subfigure}{0.48\linewidth}
        \centering
        \includegraphics[width=\linewidth]{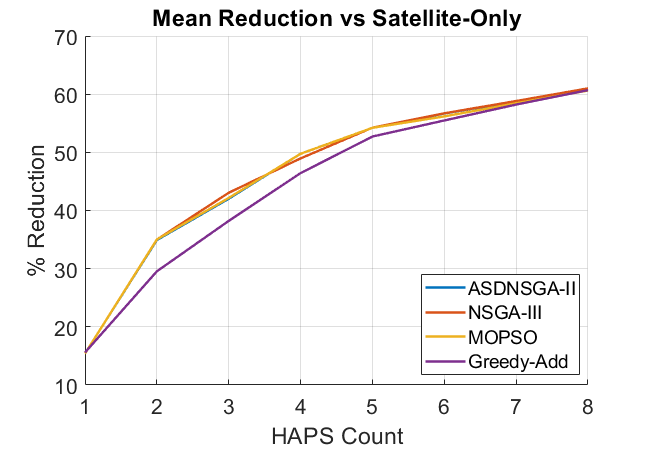}
        \caption{Mean reduction.}
    \end{subfigure}
    \hfill
    \begin{subfigure}{0.48\linewidth}
        \centering
        \includegraphics[width=\linewidth]{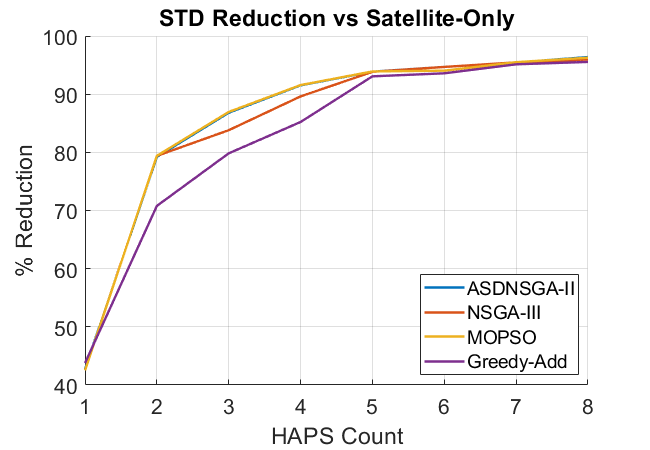}
        \caption{STD reduction.}
    \end{subfigure}

    \vspace{0.25cm}

    \begin{subfigure}{0.48\linewidth}
        \centering
        \includegraphics[width=\linewidth]{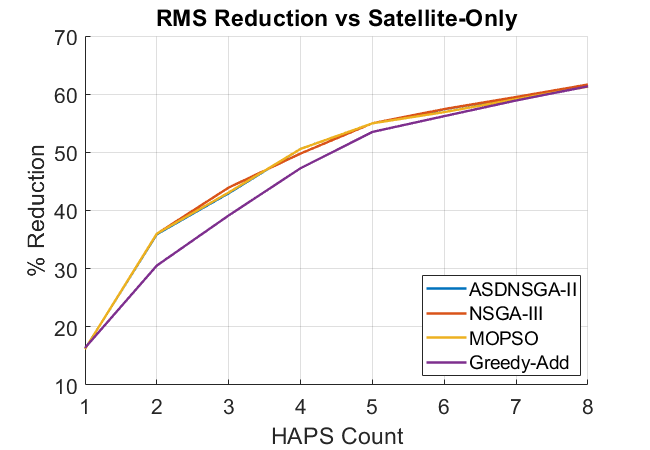}
        \caption{RMS reduction.}
    \end{subfigure}
    \hfill
    \begin{subfigure}{0.48\linewidth}
        \centering
        \includegraphics[width=\linewidth]{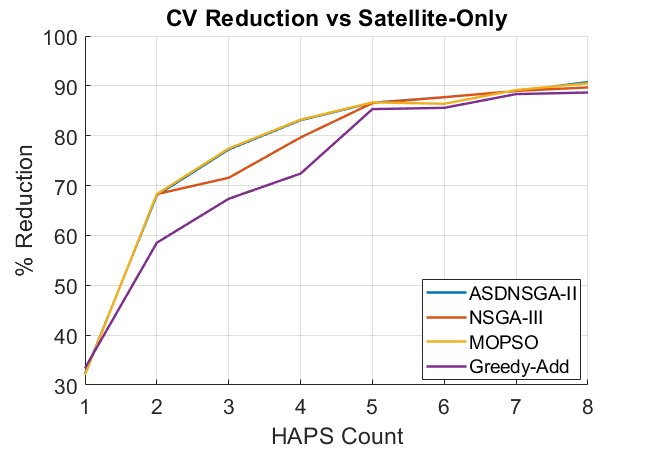}
        \caption{CV reduction.}
    \end{subfigure}

    \caption{Performance gains relative to the satellite-only baseline, shown for mean, STD, RMS, and CV of the average 3D PEB computed across snapshots over one sidereal day.}
    \label{fig:stats plot}
\end{figure}

Based on these statistics, Fig.~\ref{fig:stats plot} visualizes the performance gains relative to the satellite-only baseline. Across all metrics, a substantial improvement is observed when the HAPS count increases from 1 to 2, with reductions of 35\% in mean, 80\% in STD, 36\% in RMS, and 68\% in CV. As the HAPS count increases further, a saturation effect becomes evident, particularly in STD and CV. For example, at 5 HAPS, the reductions in mean, STD, RMS, and CV reach 54\%, 94\%, 55\%, and 87\%, respectively, relative to the satellite-only case. These results demonstrate that the proposed framework, along with the three metaheuristic algorithms, effectively improves not only the worst-case geometric conditions but also the overall robustness of the system, as evidenced by the pronounced reductions in STD and CV. Moreover, the results indicate that adding more than five HAPS contributes only marginal additional benefit, reflecting the diminishing returns associated with increased coverage overlap and geometric redundancy.

Interestingly, at 5 HAPS, the Greedy-Add algorithm achieves performance comparable to that of the metaheuristic methods. This is because, as seen in Fig.~\ref{fig:skyplot (5-haps)}, the five greedily selected HAPS positions happen to form a well-distributed geometry across the ROI, thereby mitigating the limitations of its sequential, non-revisiting selection strategy.

\begin{figure}[!t]
\centering
\includegraphics[width=0.9\columnwidth]{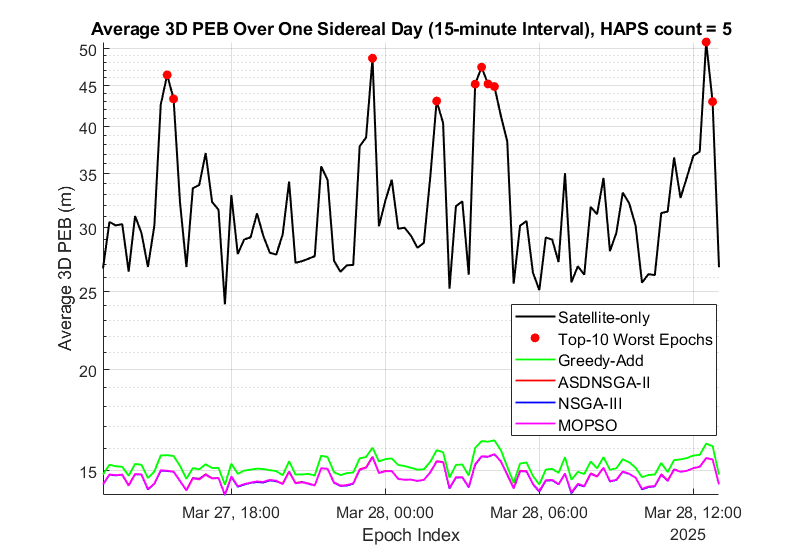}
\caption{Average 3D PEB over one sidereal day at 15-minute intervals for the 5-HAPS configuration.}
\label{fig:avg. 3d peb over one sidereal day}
\end{figure}

To illustrate the system performance using the optimally identified HAPS configurations, Fig.~\ref{fig:avg. 3d peb over one sidereal day} shows the average 3D PEB over one sidereal day based on the optimal 5-HAPS placements obtained from each algorithm. The figure also highlights the 10 snapshots with the highest average 3D PEB. The results demonstrate a substantial reduction in average 3D PEB compared with the satellite-only case, and the absence of pronounced spikes across the day indicates strong robustness under varying satellite geometries. These results confirm both the effectiveness and efficiency of using only the top 10 worst-PEB snapshots in constructing the objective function. This reduced snapshot set substantially lowers computational cost while still steering the algorithms toward solutions that resolve the most challenging geometric conditions and deliver consistently strong performance across the full sidereal day.

\subsection{GDOP Analysis}

\begin{figure}[!t]
\centering
\includegraphics[width=0.9\columnwidth]{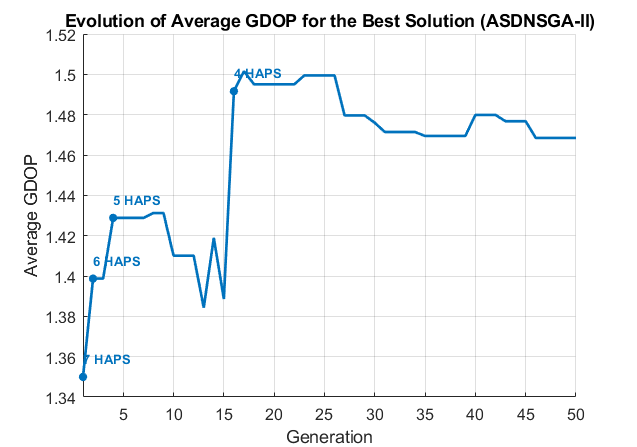}
\caption{Evolution of the average GDOP in the best solution of ASDNSGA-II across generations.}
\label{fig:gdop evolution}
\end{figure}

To show that a lower GDOP does not necessarily correspond to better system performance, Fig.~\ref{fig:gdop evolution} illustrates the evolution of the average GDOP in the best solution identified by ASDNSGA-II over successive generations. As expected, GDOP increases when the HAPS count is reduced. For example, a clear rise appears when moving from five to four HAPS. However, GDOP also fluctuates even when the HAPS count remains unchanged, reflecting its strong dependence on the spatial configuration of HAPS and environment rather than their number alone; this is particularly evident in the 4-HAPS and 5-HAPS cases.

More importantly, with the HAPS count held constant, the best solution can yield the same or lower average 3D PEB, yet the corresponding GDOP may still rise. For example, at generation 14, GDOP rises noticeably despite the HAPS count remaining at five. This discrepancy highlights that GDOP-based placement alone fails to capture environment-dependent factors, such as urban blockages and LOS/NLOS variations, and is therefore insufficient as a standalone performance metric in realistic deployment scenarios.

% \begin{figure}[!t]
% \centering
% \includegraphics[width=0.9\columnwidth]{images/skyplot, gps satellites, wallstreet.png}
% \caption{Skyplot illustrating satellite locations at sampled snapshots throughout one sidereal day.}
% \label{fig:skyplot_satellites}
% \end{figure}

\subsection{Runtime Comparison and Bottleneck Analysis}
\begin{figure}[!t]
\centering
\includegraphics[width=0.9\columnwidth]{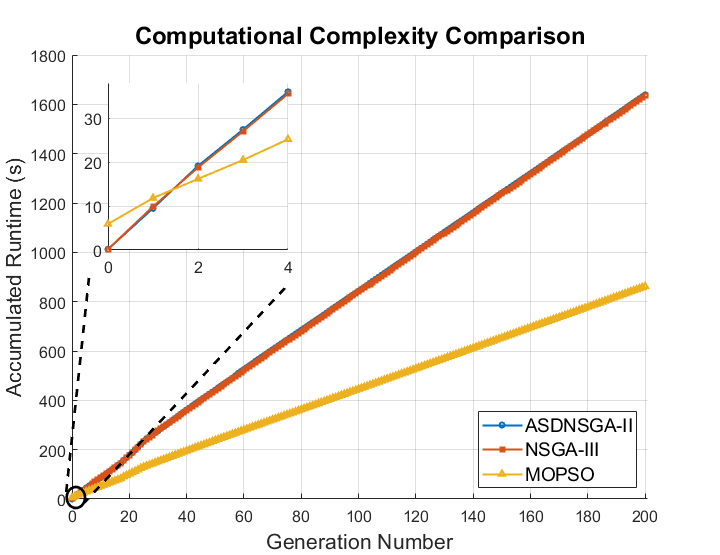}
\caption{Runtime performance comparison among ASDNSGA-II, NSGA-III, and MOPSO in the Wall Street scenario.}
\label{fig:runtime}
\end{figure}

To compare runtime performance and analyze computational bottlenecks, we present the runtime behavior of ASDNSGA-II, NSGA-III, and MOPSO across generations for the Wall Street scenario, as shown in Fig.~\ref{fig:runtime}. We can see that all three algorithms exhibit an approximately linear increase in accumulated runtime with respect to the number of generations, indicating a near-constant per-generation computational cost. Notably, MOPSO shows a higher initial runtime offset, primarily due to the overhead associated with initializing the particle swarm and constructing the external archive for leader selection, where FNS is employed. However, after initialization, MOPSO demonstrates more efficient runtime scaling than the NSGA-based methods, as FNS is applied less frequently and typically on a smaller set of solutions during archive updates. Consequently, its per-generation computational cost is lower, with the main overhead arising from leader selection and particle updates.

In contrast, ASDNSGA-II and NSGA-III rely on frequent applications of FNS at every generation on the full population, resulting in higher computational overhead due to the quadratic complexity of pairwise dominance comparisons. In addition, ASDNSGA-II incurs extra cost from variable-length solution handling via the ANND-based matching process, while NSGA-III requires associating individuals with predefined reference directions, introducing additional distance computations in the objective space. As a result, these two GAs exhibit comparable runtime behavior.

From a scalability perspective, the computational burden of all methods increases with the size of ROI and the resolution of the candidate HAPS grid. In particular, the greedy baseline approach, which evaluates a large set of candidate positions directly, scales poorly as the grid becomes finer, since its complexity is dominated by exhaustive candidate evaluation; in our experiments, it requires approximately 26 hours to complete on the aforementioned 192-core HPC cluster for the Wall Street scenario. Evolutionary approaches, while incurring higher per-generation costs, are less sensitive to grid density as they explore the solution space stochastically. Nevertheless, their runtime still increases with problem size due to more expensive objective evaluations and potentially larger population sizes required to maintain solution diversity. 

To scale to larger ROIs and/or finer HAPS grids, higher-performance computing resources can be leveraged, and the contribution of individual operators to solution quality can be analyzed to eliminate those with limited impact. Overall, the three metaheuristic algorithms consistently match or exceed the solution quality of the greedy baseline while providing a substantial leap in runtime efficiency.

\subsection{Generalization Across 3D City Models}
\begin{figure*}[t]
\centering
\begin{subfigure}[t]{0.32\textwidth}
    \centering
    \includegraphics[width=\linewidth]{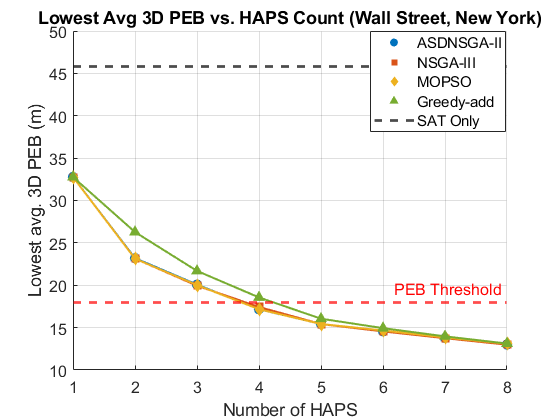}
    \caption{Wall Street.}
    \label{fig:final_wallstreet}
\end{subfigure}
\hfill
\begin{subfigure}[t]{0.32\textwidth}
    \centering
    \includegraphics[width=\linewidth]{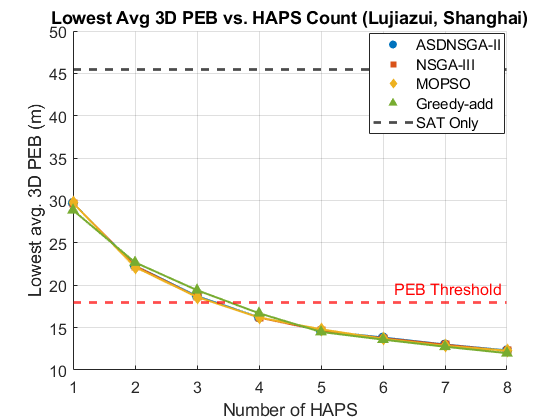}
    \caption{Lujiazui.}
    \label{fig:final_lujiazui}
\end{subfigure}
\hfill
\begin{subfigure}[t]{0.32\textwidth}
    \centering
    \includegraphics[width=\linewidth]{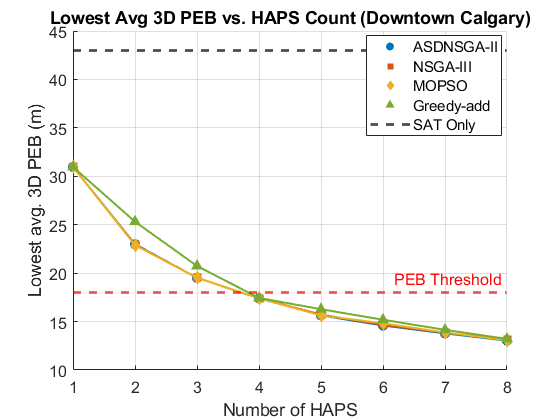}
    \caption{Downtown Calgary.}
    \label{fig:final_calgary}
\end{subfigure}
\caption{Final-generation Pareto fronts of average 3D PEB versus HAPS count for different algorithms across three urban regions: Wall Street (New York City), Lujiazui (Shanghai), and downtown Calgary.}
\label{fig:lowest avg 3d peb vs haps count_final}
\end{figure*}

The proposed metaheuristic framework can be generalized to other city models. In this work, we evaluate its generalizability across three representative city models: Wall Street (New York City), Lujiazui (Shanghai), and downtown Calgary. Fig.~\ref{fig:lowest avg 3d peb vs haps count_final} shows the final-generation Pareto fronts of average 3D PEB versus HAPS count for the three metaheuristic algorithms--ASDNSGA-II, NSGA-III, and MOPSO--along with the Greedy-Add baseline. As a reference, the average 3D PEB for the satellite-only case is also included. As shown, the metaheuristic algorithms exhibit nearly identical performance across all HAPS counts and urban scenarios. This behavior can be attributed to two main factors: (i) the enforced methodological consistency, including shared operators and diversity mechanisms, which aligns their search behavior; and (ii) the structure of the optimization problem, where a limited number of high-quality configurations exist and geometric saturation reduces sensitivity to algorithmic differences. Consequently, despite their different search mechanisms, the algorithms are similarly effective in identifying near-optimal solutions. Moreover, the relative performance compared to the Greedy-Add baseline depends on the HAPS count, with the most noticeable differences occurring at low-to-moderate HAPS regime. In this region, the metaheuristic algorithms occasionally outperform the Greedy-Add approach, while their performance is largely comparable at higher HAPS counts.

% The metaheuristic algorithms (ASDNSGA-II, NSGA-III, and MOPSO) produce broadly similar spatial distributions, reflecting their nearly identical PEB performance seen in Fig.~\ref{fig:lowest avg 3d peb vs haps count_final}. In contrast, the configuration obtained by Greedy-Add deviates more noticeably from the metaheuristic solutions. This difference stems from its sequential, non-revisiting selection procedure, which prevents the algorithm from repositioning earlier choices even when additional HAPS are introduced, thereby limiting its flexibility compared to population-based search methods.

For the 1-HAPS case, all three metaheuristic algorithms manage to identify a solution with performance comparable to the greedy baseline. This is expected and serves as an implementation check: Greedy-Add is essentially a fingerprinting-based method that evaluates the full candidate set at the first step, and thus it should identify the optimal single-HAPS placement. However, we observe that for Lujiazui, the performance of all metaheuristic algorithms is slightly worse than the greedy method. This can be attributed to the spatial distribution of the poor-visibility receivers, which causes the region center, defined by the geodetic centroid, to deviate from the center of the area covered by these receivers. Since the metaheuristic framework leverages the region center in candidate generation and feasibility constraints, this mismatch may bias the search away from the most beneficial HAPS placement for the targeted receivers. In contrast, the greedy approach directly evaluates all candidate locations based on the receiver-dependent objective and is therefore less sensitive to this deviation, allowing it to more reliably identify the optimal placement in this scenario. It is worth noting that the use of a region center is a simplifying assumption, and more refined strategies could mitigate this issue. For example, a more representative definition of the region center could be adopted, the skyline elevation threshold could be increased to capture more receivers with poor visibility, or, if computational resources permit, the reliance on a region center could be removed entirely.

For HAPS counts between 2 and 6, all three metaheuristic algorithms consistently outperform the Greedy-Add method. This behavior is intuitive. Greedy-Add selects HAPS sequentially based solely on the largest immediate marginal improvement, without revisiting earlier choices. In contrast, the optimal multi-HAPS configuration depends on how the candidate HAPS collectively complement satellite geometries and the receiver environments. As a result, a globally optimal configuration for a higher HAPS count may require placements that differ substantially from those that appear optimal at lower counts. Due to the deviation of the region center in Lujiazui, the performance advantage of the metaheuristic algorithms is less pronounced compared to the other two cases.

\begin{figure*}[t]
\centering
\begin{subfigure}[t]{0.32\textwidth}
    \centering
    \includegraphics[width=\linewidth]{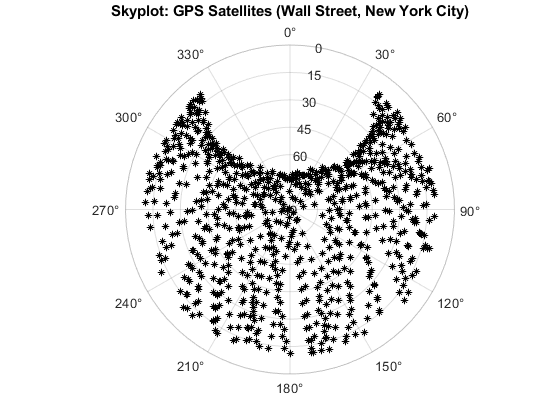}
    \caption{Wall Street.}
    \label{fig:skyplot_wallstreet}
\end{subfigure}
\hfill
\begin{subfigure}[t]{0.32\textwidth}
    \centering
    \includegraphics[width=\linewidth]{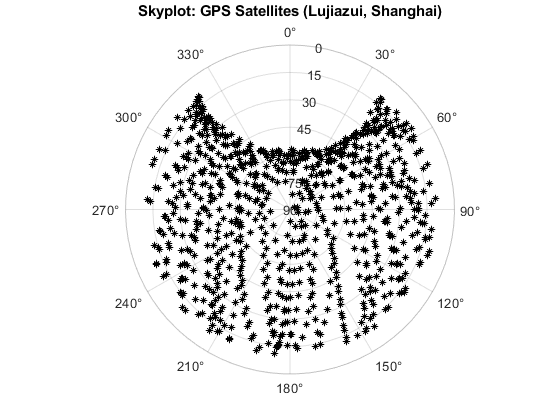}
    \caption{Lujiazui.}
    \label{fig:skyplot_lujiazui}
\end{subfigure}
\hfill
\begin{subfigure}[t]{0.32\textwidth}
    \centering
    \includegraphics[width=\linewidth]{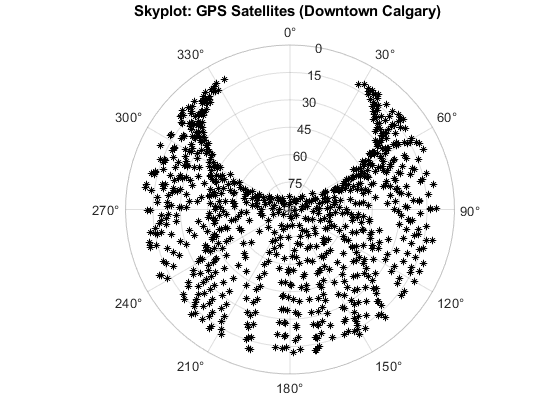}
    \caption{Downtown Calgary.}
    \label{fig:skyplot_calgary}
\end{subfigure}
\caption{Skyplots of satellite locations over one sidereal day for Wall Street (New York City), Lujiazui (Shanghai), and downtown Calgary.}
\label{fig:skyplot_satellites}
\end{figure*}

\begin{figure*}[t]
\centering
\begin{subfigure}[t]{0.32\textwidth}
    \centering
    \includegraphics[width=\linewidth]{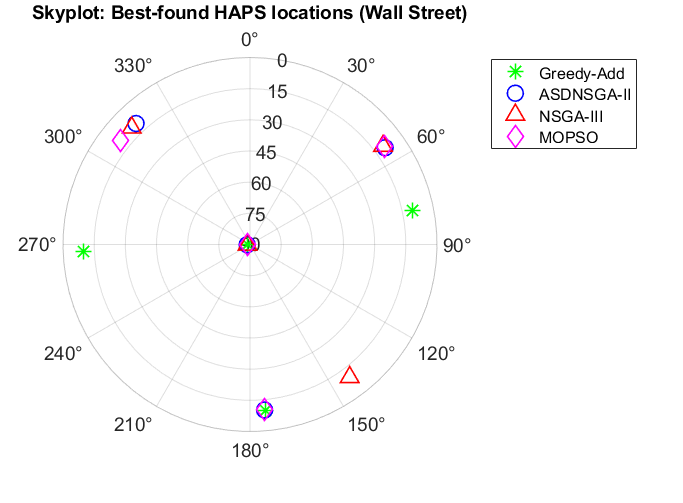}
    \caption{Wall Street.}
    \label{fig:skyplot_wallstreet}
\end{subfigure}
\hfill
\begin{subfigure}[t]{0.32\textwidth}
    \centering
    \includegraphics[width=\linewidth]{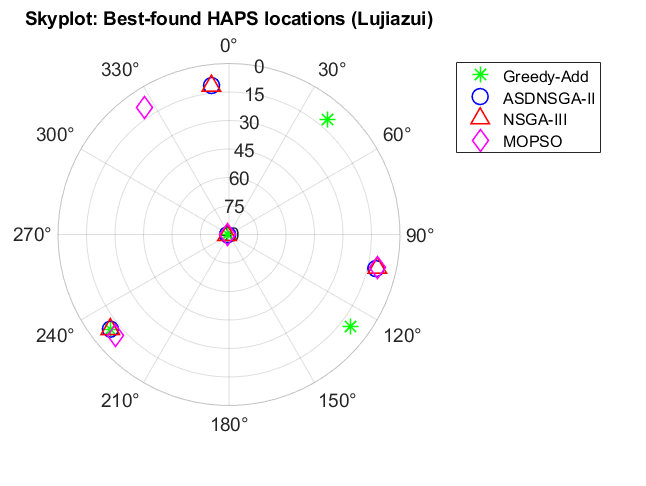}
    \caption{Lujiazui.}
    \label{fig:skyplot_lujiazui}
\end{subfigure}
\hfill
\begin{subfigure}[t]{0.32\textwidth}
    \centering
    \includegraphics[width=\linewidth]{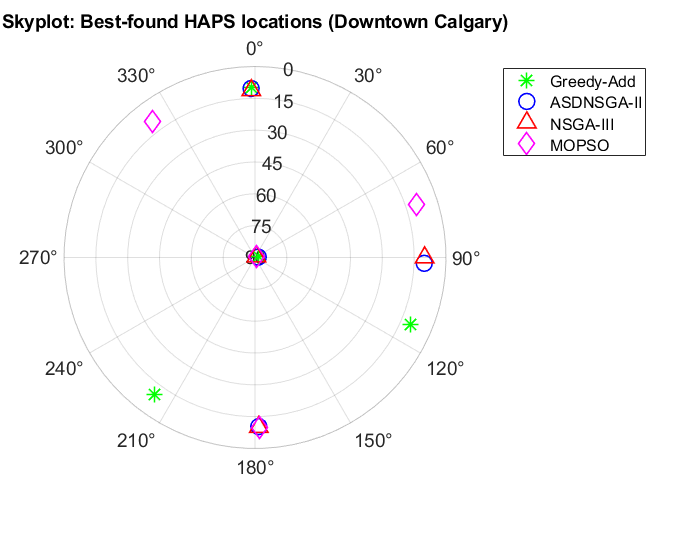}
    \caption{Downtown Calgary.}
    \label{fig:skyplot_calgary}
\end{subfigure}
\caption{Skyplots of optimal 4-HAPS configurations for Wall Street (New York City), Lujiazui (Shanghai), and downtown Calgary.}
\label{fig:skyplot_cities}
\end{figure*}

Interestingly, the performance of the metaheuristic algorithms for the 4-HAPS configuration in Calgary is comparable to the greedy baseline. This can be attributed to the inherent limitations in GPS satellite geometry, particularly at higher latitudes. As illustrated in Fig.~\ref{fig:skyplot_satellites}, the skyplots of GPS satellites across sampled snapshots across the three urban regions reveal characteristic arc-shaped trajectories resulting from the $55^\circ$ inclined GPS orbits. This leads to non-uniform sky coverage across these regions. In particular, the azimuth sector between approximately $330^\circ$ and $30^\circ$ exhibits noticeably fewer satellite appearances. This distribution is consistent with expected GPS visibility patterns at mid-northern latitudes and reflects natural constellation dynamics rather than a simulation artifact. Furthermore, the Calgary skyplot shows reduced high-elevation satellite visibility and a stronger concentration of satellites at lower elevations, which is a direct consequence of the constellation’s orbital inclination. 

This effect is further reflected in the skyplots of the optimal 4-HAPS placements. As we can see from Fig.~\ref{fig:skyplot_cities}, for Wall Street and Lujiazui, where GPS coverage is more evenly distributed compared to Calgary, the Greedy-Add method tends to place HAPS sequentially in directions that provide the largest immediate marginal improvement, often resulting in placements along roughly orthogonal directions with one HAPS at high elevation. However, in the 4-HAPS case, this sequential strategy can lead to uneven azimuthal coverage, leaving certain sectors underrepresented. In contrast, the metaheuristic algorithms jointly optimize all HAPS positions and achieve more uniformly distributed configurations with better azimuth diversity. As a result, all metaheuristic algorithms attain lower average 3D PEB than the greedy baseline. In contrast, for Calgary, the skyplot reveals a noticeable lack of satellite coverage in the northern direction. Under such constrained geometry, the benefit of achieving uniform azimuthal diversity is reduced. Consequently, the best 4-HAPS configuration identified by the Greedy-Add method may exhibit azimuth diversity comparable to that obtained by the metaheuristic algorithms, leading to similar performance.

At higher HAPS counts, the performance gap narrows and the Greedy-Add method becomes competitive with the metaheuristics, as shown in Fig.~\ref{fig:lowest avg 3d peb vs haps count_final}. Under the prescribed simulation setup, the results indicate that four HAPS are adequate to reduce the average 3D PEB below the chosen threshold across all three urban regions. However, it is worth noting that this threshold serves only as a practical decision rule and has a negligible effect on the optimal configurations identified. For instance, if the threshold were tightened to 15 m, the minimum HAPS count required would increase to six for Wall Street and downtown Calgary, and to five for Lujiazui.

\section{Conclusion}
This paper proposes a metaheuristic framework for jointly optimizing the number and placement of HAPS to enhance GNSS-based localization in dense urban environments. By integrating high-fidelity 3D city models, ray-tracing-based LOS/NLOS classification, and multi-objective optimization, the framework effectively addresses the non-convex and discrete nature of the problem. Across diverse urban conditions modeled through different GMM parameter settings, the results demonstrate strong robustness of the proposed approach, with all three algorithms consistently achieving high-quality Pareto fronts that outperform the greedy baseline, particularly in the low-to-moderate HAPS regime. For the considered dense urban geometries, we show that four HAPS are sufficient to satisfy the 18-m average 3D PEB threshold, while configurations with two to six HAPS provide the most significant gains in both accuracy and robustness. Within this range, the metaheuristic methods outperform the greedy baseline by better capturing the joint placement dependencies, whereas performance differences diminish at higher HAPS counts due to geometric saturation.

Among the three metaheuristic algorithms, we show that all exhibit efficient convergence within the first 50 generations, confirming their suitability for large-scale, non-convex optimization. MOPSO, in particular, demonstrates a favorable balance between solution quality and computational efficiency, achieving performance comparable to the genetic algorithm-based methods while incurring lower runtime overhead. In addition, the framework generalizes well across different urban environments, including regions with varying building morphology and satellite visibility conditions, indicating its adaptability beyond the primary case study. System-level insights further reveal that optimal configurations consistently include a near-zenith HAPS combined with azimuthally distributed platforms to enhance geometric diversity. Moreover, the results highlight that traditional metrics such as GDOP are insufficient in urban settings, as they fail to capture environment-dependent effects such as blockage and multipath. Overall, the proposed framework provides an effective and scalable solution for designing HAPS-assisted localization systems under realistic urban conditions.

Several directions remain for future work to further enhance the practicality and generality of the framework. For instance, extending from a GPS-only scenario to multi-constellation GNSS (e.g., GPS combined with Galileo or GLONASS) could improve satellite availability and potentially reduce the number of required HAPS. Incorporating dynamic receiver trajectories, such as vehicles moving through urban environments, would enable evaluation under time-varying LOS/NLOS conditions and provide insight into real-time performance. In addition, investigating the impact of signal bandwidth and frequency bands (e.g., L-band versus Ka-band) would offer a deeper understanding of propagation effects on localization accuracy in urban settings. For more realistic and deployment-ready solutions, future work should also consider practical system constraints, including HAPS-to-HAPS interference, power budget limitations, platform reliability and failure scenarios, and the feasibility of maintaining desired stratospheric positions under wind dynamics. Integrating these factors into the optimization framework would further improve its applicability to real-world deployments.

\begin{IEEEbiography}[{\includegraphics[width=1in,height=1.25in,clip,keepaspectratio]{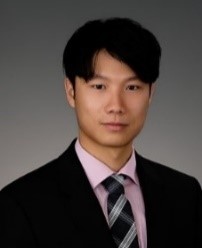}}]{Hongzhao Zheng}
(Member, IEEE) received the B.Eng. (Hons.) degree in Engineering Physics in 2019 and the Ph.D. degree in Electrical and Computer Engineering in 2025, both from Carleton University, Ottawa, ON, Canada. His thesis topic is a hybrid positioning system in a non-terrestrial network for urban areas. He is currently a Research Associate in the Carleton Non-Terrestrial Networks (Carleton-NTN) Lab.

Dr. Zheng received the best paper award at IEEE WiSEE 2022, the student travel grant at IEEE ICC 2024, and the senate medal for outstanding academic achievement at the doctoral level.
\end{IEEEbiography}

\begin{IEEEbiography}[{\includegraphics[width=1in,height=1.25in,clip,keepaspectratio]{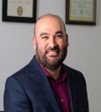}}]{Mohamed Atia} (Senior Member, IEEE) received the B.Sc. and M.Sc. degrees in computer systems from Ain Shams University, Cairo, Egypt, in 2000 and 2006, respectively, and the Ph.D. degree in electrical and computer engineering from Queen's University, Kingston, ON, Canada, in 2013. He is currently an Associate Professor in the Department of Systems and Computer Engineering at Carleton University, Ottawa, ON, Canada, where he founded the Embedded and Multi-Sensor Systems Lab (EMSLab). Before joining academia, he spent several years in the industry developing advanced algorithmic systems that integrate artificial intelligence, machine learning, signal processing, and estimation techniques for applications including natural language processing, speech recognition, and multi-sensor navigation. Dr. Atia's research interests include real-time embedded software systems, autonomous navigation, sensor fusion, simultaneous localization and mapping (SLAM), positioning in GNSS-denied environments, and intelligent robotic perception. He has authored or coauthored over 100 peer-reviewed publications in journals, conferences, and books, and holds five granted patents.

\end{IEEEbiography}

\begin{IEEEbiography}[{\includegraphics[width=1in,height=1.25in,clip,keepaspectratio]{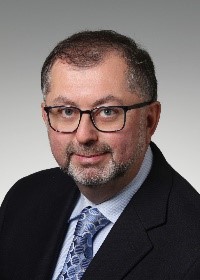}}]{Halim Yanikomeroglu}
(Fellow, IEEE) is a Chancellor's Professor in the Department of Systems and Computer Engineering at Carleton University, Canada; he is also the Director of Carleton-NTN (Non-Terrestrial Networks) Lab. He is a Fellow of IEEE, Engineering Institute of Canada (EIC), Canadian Academy of Engineering (CAE), and Asia-Pacific Artificial Intelligence Association (AAIA). Dr. Yanikomeroglu has coauthored a high number of papers in 33 different IEEE journals; he also has 43 granted patents. He has supervised or hosted at Carleton 170+ postgraduate researchers; several of his former team members have become professors in Canada, US, UK, and around the world. He gives around 20 invited seminars, keynotes, panel talks, and tutorials every year. He has served as the Steering Committee Chair, General Chair, and Technical Program Chair of several major international IEEE conferences, as well as in the editorial boards of several IEEE periodicals. He also served as a Distinguished Speaker for IEEE Communications Society and IEEE Vehicular Technology Society, and an Expert Panelist of the Council of Canadian Academies (CCA|CAC). Dr. Yanikomeroglu received many awards for his research, teaching, and service. He holds a PhD degree in electrical and computer engineering from University of Toronto.
\end{IEEEbiography}

\vfill

\end{document}